\documentclass[useAMS,usenatbib,usegraphicx, letterpaper]{mn2e}

\usepackage{color}
\def\eref#1{equation (\ref{eq:#1})}
\def\eeref#1{(\ref{eq:#1})}

\title{Three-dimensional hydrodynamic simulations of asymmetric pulsar
  wind bow shocks}

\author[M. Vigelius et al.]{M.~Vigelius, $^1$ A.~Melatos, $^1$
 S. Chatterjee, $^{2,3}$ B. M. ~Gaensler
 $^{3,4,5}$ and P.~Ghavamian$^6$ \\
 $^1$ School of Physics, University of Melbourne, Parkville, VIC
 3010, Australia; mvigeliu@physics.unimelb.edu.au \\
 $^2$ Jansky Fellow, National Radio Astronomy Observatory, PO Box O,
 Socorro, NM 87801 \\
 $^3$ Harvard-Smithsonian Center for Astrophysics,60 Garden Street
 , Cambridge, MA 02138, USA \\
 $^4$ Alfred P. Sloan Research Fellow \\
 $^5$ Present address: School of Physics A29, University of Sydney,
 NSW 2006, Australia \\
 $^6$ Department of Physics and Astronomy, Johns Hopkins University,
 3400 North Charles Street, Baltimore, MD 21218-2686, USA}

\begin{document}

\date{Accepted 2006 October 13. Received 2006 October 13; in original form 2006 July 03}

\maketitle

\begin{abstract}
  We present three-dimensional, nonrelativistic, hydrodynamic
  simulations of bow shocks in pulsar wind nebulae. The simulations
  are performed for a
  range of initial and boundary conditions to quantify the degree of
  asymmetry produced by latitudinal variations in the momentum flux of
  the pulsar wind, radiative cooling in the postshock flow, and
  density gradients in the interstellar medium (ISM). We find
  that the bow shock is stable even when travelling through a strong ISM
  gradient.  We demonstrate how the shape of the bow shock changes when the pulsar
  encounters density variations in the ISM. We show that a density wall can
  account for the peculiar bow shock shapes of the nebulae around
  PSR J2124-3358 and PSR B0740-28.  A wall produces kinks in the
  shock, whereas a smooth ISM density gradient tilts the shock.
  We conclude that the anisotropy of the wind momentum flux alone
  cannot explain the observed bow shock morphologies but it is instead
  necessary to take into account external effects.
  We show that the analytic (single layer, thin shell)
  solution is a good approximation when the momentum flux is
  anisotropic, fails for a steep ISM density
  gradient, and approaches the numerical
  solution for efficient cooling. We provide analytic expressions for the latitudinal
  dependence of a vacuum-dipole wind and the associated shock
  shape, and compare the results to a split-monopole wind. We find
  that we are unable to distinguish between these two wind models
  purely from the bow shock morphology.
\end{abstract}

\bibliographystyle{mn2e}


\section{Introduction}
After a pulsar escapes its supernova remnant, it typically travels
supersonically through the interstellar
medium (ISM), sometimes with speeds in excess of 1000 km s$^{-1}$
\citep{Chatt05}. At the same time, pulsars emit highly relativistic
winds, probably consisting of electron-positron pairs,
with bulk Lorentz factor between $10^2$ and $10^6$ \citep{Kennel84,Spitkovsky04}. The
interaction of the pulsar wind with the ISM produces a
characteristic multilayer shock structure \citep{Bucc02,Swaluw03}.

The multilayer structure of pulsar wind nebulae (PWN) gives rise to a
variety of observable features at different wavelengths;
see \citet{Gaensler06a} for a recent review. The outer
bow shock excites neutral hydrogen atoms either collisionally or by
charge exchange, which then de-excite and
emit optical H$\alpha$ radiation \citep{Bucc01}. The microphysics of
these excitation processes was treated in the context of supernova
remnants by \citet{Chev78} and \citet{Chev80}, while
\citet{Ghav01} performed detailed calculations of the
ionization structure and optical spectra produced by non-radiative
supernova remnants in partially neutral gas. Presently, six PWN have
been detected in the optical band \citep{Gaensler06a, Kaspi04}, namely around the
pulsars PSR B1957+20 \citep{Kulkarni88}, PSR B2224+65 \citep{Cordes93},
also called the Guitar Nebula, PSR B0740$-$28 \citep{Jones02}, PSR J0437$-$4715
\citep{Bell95}, RX J1856.5$-$3754 \citep{Kerkwijk01}, and PSR J2124$-$3358 \citep{Gaensler02}.

Pulsar bow shocks directly probe the conditions in the local
ISM. The size of the bow shock scales with the
standoff distance \citep{Chatt02, Wilkin96}, defined as the point
where the ISM ram pressure balances the wind ram pressure. As the wind ram pressure is given
directly by the (measurable) spin-down luminosity of the pulsar for
both young pulsars \citep{Kennel84} and recycled millisecond
pulsars \citep{Stappers03}, the size of the bow shock can put an upper
limit on the ISM density, given a measurement of the pulsar's proper motion.
Furthermore, the morphology of the shock gives detailed information about the
inhomogeneity of the ISM. Density gradients have been proposed to explain the
peculiar shapes of the bow shocks around the pulsars PSR B0740$-$28
\citep{Jones02} and PSR J2124$-$3358 \citep{Gaensler02}. 
\citet{Chatterjee06} recently showed that the kink in the PWN around
PSR J2124$-$3358 can be explained if the pulsar is travelling through a
density discontinuity in the ISM (a ``wall'').

Our theoretical understanding of the wind acceleration mechanism,
in particular the angular distribution of its momentum flux, is still
incomplete. \citet{Goldreich69} showed that a pulsar has an extended
magnetosphere filled with charged particles that are accelerated
into a surrounding wave zone \citep{Coroniti90}. Although the nature of the energy transport in the
pulsar wind is still an unsolved question, probably governed by
non-ideal magnetohydrodynamics \citep{Melatos96}, all models agree
that it is dominated by the Poynting flux near the star. However,
observations of the Crab nebula \citep{Kennel84} and PSR B1509$-$58
\citep{Gaensler02a} imply a value for the magnetisation
parameter, $\sigma$, defined as the ratio of the Poynting flux to the
kinetic energy flux, satisfying $\sigma < 0.01$ at the wind termination shock.
In other words, a transition from high $\sigma$ to low $\sigma$ must occur between
the magnetosphere and the termination shock, a problem
commonly referred to as the $\sigma$ paradox. While it is generally
assumed that the angular dependence of the momentum flux is preserved
during this transition \citep{DelZanna04}, the exact dependence is
unknown. The split-monopole model predicts a mainly equatorial flux
\citep{Bogovalov99, Bucc06} while
wave-like dipole models \citep{Melatos96} predict a more complicated distribution. The study of
the termination shock structure, including its variability
\citep{Spitkovsky04, Melatos05}, thus provides valuable
information about the angular distribution of the wind momentum
flux and hence the electrodynamics of the pulsar magnetosphere.

PWN also emit synchrotron radiation at radio and X-ray wavelengths
\citep{Gaensler06a, Kaspi04} from wind particles which are accelerated in the
inner shock and subsequently gyrate in the nebular magnetic
field \citep{Kennel84b, Spitkovsky99}. If the pulsar moves supersonically, the
(termination) shock is strongly confined by the ISM ram
pressure \citep{Bucc02,Swaluw03} and elongates opposite
the direction of motion. X-ray-emitting particles with short
synchrotron lifetimes directly probe the inner flow
structure of the PWN. With the help of combined X-ray and radio
observations, \citet{Gaensler04} identified the termination shock and
post shock flow in the PWN around PSR J1747$-$2958, the Mouse
Nebula. They found that the data are consistent with an isotropic
momentum flux distribution, a statement that holds for some
other bow shocks, e.g. PSR B2224$-$65
\citep{Chatt02}, but not all, e.g. IC433 \citep{Gaensler06a}.

Before we can use PWN to probe the ISM, a precise knowledge of the inner flow
structure and the response of the system to external influences is
important. Although \citet{Wilkin96} found an analytic solution
for the shape of the bow shock in the limit of a thin
shock, the complexity of the problem generally requires a
numerical treatment. \citet{Bucc02} performed hydrodynamic,
2.5-dimensional, cylindrically symmetric simulations to explore the
validity of a two-shell approximation to the problem
and elucidated the inner, multilayer shock
structure. \citet{Swaluw03} modelled a pulsar passing through its own
supernova remnant and found that the bow shock nebula remains
undisrupted.

Of course, the validity of a hydrodynamic approach is limited. To compute synchrotron
maps, magnetic fields must be included. Relativistic
magnetohydrodynamic (RMHD) simulations by \citet{Komissarov03,Komissarov04}
explain the peculiar jet-like emission in the Crab nebula by
magnetic collimation of the back flow of an anisotropic pulsar wind, a result
confirmed later by \citet{DelZanna04}. RMHD simulations of bow shocks
around fast moving pulsars by \citet{Bucc05} and \citet{Bogovalov05}
assume cylindrical symmetry and an isotropic pulsar wind.

The chief new contribution of this paper is to relax the assumption of
cylindrical symmetry. We perform fully three-dimensional, hydrodynamic
simulations of PWN, focusing on observable features which can be
compared directly to observations. Although pulsar winds are generally
expected to be symmetric around the pulsar rotation axis (even if the
momentum flux varies strongly with latitude), there are external factors, like an ISM density
gradient, which destroy the cylindrical symmetry. The
existence of two very different speeds, namely that of the
relativistic wind and the pulsar proper motion, presents a significant numerical
challenge in three dimensions, which we overcome by using a parallelised hydrodynamic
code, with adaptive-mesh Godunov-type shock-capturing capabilities.

After describing the numerical method in
section \ref{sec:num_method}, we present the results of our
simulations for different types of asymmetry, namely anisotropic momentum flux
(section \ref{sec:anisotropy}), the effects of uniform cooling (section
\ref{sec:cooling}), an ISM density gradient (section
\ref{sec:densgrad}), and an ISM barrier or ``wall'' (section
\ref{sec:wall}). We conclude by discussing what information can be
deduced about the ISM and the pulsar wind when the results are combined
with observations in section \ref{sec:discussion}.

\section{Numerical Method}
\label{sec:num_method}

\subsection{Hydrodynamic model}

The simulations are performed using {\sc Flash}
\citep{Fryxell00}, a parallelized hydrodynamic solver based on the
second-order piecewise parabolic method (PPM).
\textsc{Flash} solves the inviscid hydrodynamic equations in
conservative form,

\begin{eqnarray}
 0 & = & \frac{\partial\rho}{\partial
  t}+\nabla\cdot\left(\rho\bmath{v}\right), \\
 0 & = & \frac{\partial(\rho\bmath{v})}{\partial t}+\nabla \cdot \left( \rho
  \bmath{v} \bmath{v} \right) + \nabla P, \\
  \label{eq:setup:energy}
  0 & = & \frac{\partial (\rho E)}{\partial t} + \nabla \cdot \left[
  \left( \rho E + P\right) \bmath{v} \right],
\end{eqnarray}
together with an ideal gas equation of state,
\begin{equation}
  P  =  \left( \gamma -1 \right) \rho \epsilon,
\end{equation}
where $\rho$, $P$, $\epsilon$ and $\bmath{v}$ denote the density,
pressure, internal energy per unit mass, and velocity, respectively, and
$E=\epsilon+|\bmath{v}|^2/2$ is the total energy per unit mass.
We take advantage of the multifluid capability of \textsc{Flash} to treat
the pulsar wind and the interstellar medium as fluids with different adiabatic
indices $\gamma_1$ and $\gamma_2$, respectively. The weighted average adiabatic
index $\gamma$ can then be computed from
\begin{equation}
\label{eq:def_mass_fraction}
\frac{1}{\gamma-1} = \frac{X_1}{\gamma_1-1} +\frac{X_2}{\gamma_2-1},
\end{equation}
where $X_i$ is the mass fraction of each fluid, advected according to
\begin{equation}
  \frac{\partial (\rho X_i)}{\partial t} + \nabla \cdot \left(\rho X_i
  \bmath{v} \right) = 0.
\end{equation}
We do not consider reactive flows, so there is only a small
amount of numerical mixing between the two fluids. For the problem
studied here, we find empirically that the results are qualitatively
the same for $\gamma_1=\gamma_2$ and $\gamma_1 \neq \gamma_2$ (see
section \ref{sec:shock_structure}) and therefore take
$\gamma_1=\gamma_2=5/3$ in most of our simulations, using the
multifluid capability only to trace the contact discontinuity.

\textsc{Flash} manages the adaptive mesh with the \textsc{Paramesh}
library \citep{MacNeice99}. The mesh is refined and coarsened in response to the second-order error
in the dynamical variables \citep[for details, see][]{Loehner87}.
We perform our simulations on a three-dimensional Cartesian grid with
$8\times8\times8$ cells per block and a maximum refinement level of
five nested blocks, giving a maximum resolution of $128^3$ cells.

We are limited by processing capacity to simulations lasting $\sim 15
\tau_0$, where $\tau_0$ is the time for the ISM to cross
the integration volume. Unfortunately, this is not always sufficient
to reach a genuine steady state. We occasionally encounter
ripples in the shock [e.g. Figure \ref{fig:setup:gammacomp} (right),
discussed further in section
\ref{sec:anisotropy}] which are numerical artifacts which diminish as
the simulation progresses.

\subsection{Boundary and initial conditions}
\label{sec:sim:bic}
\begin{figure}
\includegraphics[width=84mm, keepaspectratio]{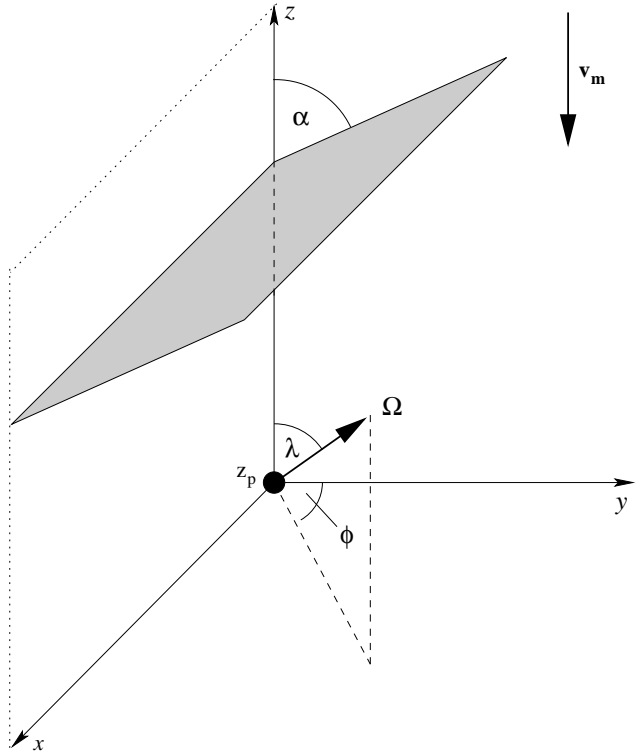}
\caption{Orientation of the pulsar, wall, and Cartesian coordinate
  axes in the pulsar's rest frame.  The pulsar is located at $(0,0,z_p$) and
  the interstellar medium moves parallel to the $z$ axis with a
  speed $v_m$.
  The symmetry axis $\mathbf{\Omega}$ of
  the wind (assumed to be the pulsar's rotation axis) is defined by
  its polar angle $\lambda$ and 
  azimuthal angle $\phi$. The wall lies perpendicular to the $y$-$z$ plane,
  making an angle $\alpha$ with respect to the $z$ axis.}
\label{fig:sim:orientation}
\end{figure}

The simulations are performed in the pulsar rest frame.
Figure \ref{fig:sim:orientation} depicts the pulsar and the orientation
of its rotation axis $\mathbf{\Omega}$ with respect to the velocity
vector $\mathbf{v}_m$ of the ISM in this frame. The momentum flux of
the pulsar wind is assumed to be cylindrically symmetric about
$\mathbf{\Omega}$, in keeping with simplified theoretical models like
the split monopole \citep{Bogovalov99, Komissarov03} or vacuum dipole
\citep{Melatos96, Melatos97, Melatos02}.

In the pulsar's rest frame, the ISM appears as a steady wind that
enters the integration volume from the upper boundary $z=z_\rmn{max}$, with velocity
$\mathbf{v}_m=-\mathbf{v}_p=-4\times10^{7}$ cm
s$^{-1}\,\bmath{\hat{z}}$, where $\mathbf{v}_p$ is the pulsar's
velocity in the observer's frame \citep{Hobbs05},
density $\rho_m=($0.33--1.30$) \times
10^{-23}$ g cm$^{-3}$, and specific internal energy $\epsilon_m=1.18 \times
10^{12}$ erg g$^{-1}$. The sound speed
$c_m=1.15\times10^{6}$ cm s$^{-1}$ and Mach number $M_m = v_m/c_m =
35$ are typical of the parameters inferred for a variety of observed
PWN \citep{Kaspi04, Bucc05}. All boundaries except $z=z_\mathrm{max}$
act as outflow boundaries, where the values of all the dynamical
variables are equalized in the boundary cells and integration
region (zero gradient condition). These boundary conditions are
justified provided that the flow speed across the boundaries exceeds
the sound speed thus preventing a casual connection of the regions
inside and outside the computational domain. We check this assumption
\emph{a posteriori} in section \ref{sec:shock_structure}. Initially, we set $\rho=\rho_m$,
$\mathbf{v}=(0,0,-v_m)$, and $\epsilon=\epsilon_m$ in every cell.

We allow for the possibility that the pulsar travels into a density
gradient. This may be either a smooth gradient perpendicular to the
pulsar's direction of motion (section \ref{sec:densgrad}) or an extended
ridge of high density material making an angle $\alpha$ with $\mathbf{v}_m$, which we call a ``wall'', also
depicted in figure \ref{fig:sim:orientation} (section \ref{sec:wall}).
The boundary conditions corresponding to these scenarios are defined
in section \ref{sec:densgrad} and \ref{sec:wall}.

The simulation parameters for the different models are listed in table
\ref{tab:models}.

\section{Anisotropic wind}
\label{sec:anisotropy}

Although the electrodynamics of pulsar wind acceleration and
collimation remains unsolved,
observations of PWN \citep{Helfand01, Gaensler02a, Roberts03} as well as
theoretical studies of split-monopole \citep{Bogovalov99, Komissarov03,Komissarov04}
and wave-like dipole \citep{Melatos96, Melatos97, Melatos02} outflows seem to favor
anisotropic momentum and energy flux distributions, where the wind
is cylindrically symmetric about $\mathbf{\Omega}$ but varies in
strength as a function of latitude. Letting $\theta$ denote the
colatitude measured relative to $\mathbf{\Omega}$, the various models
predict the following momentum flux distributions:
$g_w (\theta) = \sin^2 \theta + 1/\sigma$ for the split monopole
\citep{Bogovalov99, Komissarov03, Arons04}, $g_w(\theta) = \cos^2 \theta+1$ for the
point dipole, and $g_w(\theta) = 2 \cos^4 \theta+6 \cos^2 \theta +5$ for the extended
dipole (\citealt{Melatos97}; see also appendix \ref{sec:app:analytic}). The
normalized momentum flux $g_w(\theta)$ is defined precisely in appendix \ref{sec:app:momflux}.

\subsection{Numerical implementation}
\label{sec:anisotropy:implementation}

We implement the wind anisotropy by varying the wind density
$\rho_w$ according to
\begin{equation}
  \label{eq:aniso:momflux}
  \rho_w(\theta)=\rho_{0,w} g_w(\theta)=\rho_{0,w}\left(c_0+c_2 \cos^p \theta \right),
\end{equation}
where $c_0$, $c_2$ and $p$ are constants, while leaving the wind velocity $\mathbf{v}_w$
constant. Strictly speaking, this is unrealistic, because it is likely
that $\mathbf{v}_w$ varies with $\theta$ too in a true pulsar
wind. However, it is expected to be an excellent approximation because
\citet{Wilkin96, Wilkin00} showed that the shape of the bow shock is a
function of the momentum flux $\rho_w v_w^2$ only, not $\rho_w$ and
$\mathbf{v_w}$ separately, at least in the thin-shell limit of a
radiative shock. We deliberately choose \eref{aniso:momflux} to have
exactly the same form as in the analytic theory \citep{Wilkin96,Wilkin00} to
assist comparison below.

In \eref{aniso:momflux}, $c_0$ and $c_2$ parametrize the
anisotropy of the wind, while $p$ can be used to model jet-like
outflows (e.g. $p \ge 4$). For $p=2$, \eref{aniso:momflux} is equivalent to equation
(109) of \citet{Wilkin00}. The split-monopole model corresponds to
$p=2$ and $c_2=-1$ \citep{Komissarov03}. Note that normalisation
requires $c_0=1-c_2/3$ for $p=2$ and $c_0=1-c_2/5$ for $p=4$.
The analytic solutions for these two cases are presented in appendix \ref{sec:app:analytic}.

Although the pulsar wind is ultrarelativistic [e.g. Lorentz factor
$10^2-10^6$ in the Crab Nebula; \citet{Kennel84, Spitkovsky04}], the shape is determined by the
ratio of the momentum fluxes alone \citep{Wilkin96, Wilkin00}. We can therefore expect our model
to accurately describe the flow structure, as long as the value of the
mean flux $\rho_{0,w} v_w^2= 4.6\times10^{-8}$ g cm$^{-1}$ s$^{-2}$ is
realistic (which it is), even though we
have $v_w \ll c$. This approach was also adopted by
\citet{Bucc02} and \citet{Swaluw03}. \citet{Bucc05} performed RMHD
simulations of PWN and a comparison of our results with theirs does not reveal any
specifically relativistic effects; RMHD simulations with
low $\sigma$ agree qualitatively with non-relativistic hydrodynamical simulations.
In other words, as long as we choose $\rho_w$ and
$\mathbf{v}_w$ such that $\rho_w v_w^2=\dot{E}/4 \pi r_w^2 c$, where
$\dot{E}$ is the pulsar's spin-down luminosity (and $r_w$ is defined
in the next paragraph), we get the same final
result.\footnote{It is important to bear in mind that this is not
  strictly true in an adiabatic (thick-shell) shock \citep{Luo90} or
  when the two winds contain different particle species \citep[see
  Appendix B of][]{Melatos95}, where $\rho_w$ and $\mathbf{v}_w$
  control the shape of the shock separately and one should use the relativistic expression
  $n_w \gamma_w m c^2 \mathbf{v}_w$ for the momentum flux, where $n_w$
is the number density, $\gamma_w$ is the Lorentz factor, and $m$ is the
particle mass, as well as a relativistic code.}

The pulsar wind has density $\rho_{0,w}=4.6\times 10^{-24}$ g
cm$^{-3}$, specific internal energy $\epsilon_w=5.9 \times 10^{12}$ erg
g$^{-1}$, and velocity $\bmath{v}_w=v_w \bmath{e}_r$, where $\bmath{e}_r$ is the
radial unit vector and $v_w=10^8$ cm s$^{-1}$. This particular
choice gives a sound speed $c_w=2.5\times 10^{6}$ cm s$^{-1}$ and Mach
number $M_w=40$. The pulsar wind is launched by restoring the dynamical variables
$\rho_w, \mathbf{v}_w$ and $\epsilon_w$ in a spherical region with radius
$r_w=6.2\times10^{15}$ cm around
$\bmath{r}_p=(0,0,z_p)$ (figure \ref{fig:sim:orientation}) to their
initial values after every timestep.

\begin{table*}
  \centering
  \caption{Model parameters. The wind momentum flux is proportional to
    $1-c_2/3+c_2\cos^p \theta$, where $\theta$ is the colatitude measured with
    respect to the pulsar spin axis $\mathbf{\Omega}$, and $\lambda$ and $\phi$ define
    the orientation of this axis (figure \ref{fig:sim:orientation} and
    section \ref{sec:anisotropy}). $R_0$ is the standoff distance as
    measured from the simulation output, $f$ stands for the cooling
    efficiency (section \ref{sec:cooling}), and $H$ is the scale-height
    of the ambient density gradient (section \ref{sec:densgrad}). The
    wall (section \ref{sec:wall}) is described by the density contrast
    $\eta=\rho_w/\rho_m$, the angle $\alpha$, and the distance over
    which the density rises ($\Delta_r$) and declines ($\Delta_d$).
    The dots indicate parameters that are not used in that simulation.}
  \begin{tabular}{@{}ccccccccccccc}
    \hline
    Model & $p$ & $c_2$ & $\lambda$ [\degr] & $\phi$ [\degr]  &
    $R_0$ [cm/$10^{16}$] & $\log f$ & $H/R_0$ & $\rho_m$ [g
    cm$^{-3}/10^{-23}$] & $\eta$ & $\alpha$ [\degr] &
    $\Delta_r/R_0$ & $\Delta_d/R_0$ \\
    \hline
    A & 2 & 3 & 45 & 0 & 1.5 & \ldots & \ldots & 1.30 & \ldots &
    \ldots & \ldots \\
    B & 2 & 3 & 90 & 0 & 1.5 & \ldots & \ldots & 1.30 & \ldots & \ldots &
    \ldots & \ldots \\
    C & 4 & 3 & 45 & 0 & 2 & \ldots & \ldots & 1.30 & \ldots & 
    \ldots & \ldots \\
    \hline
    A1 & 2 & --1 & 45 & 0 & 2.2 & \ldots & \ldots & 1.30 & \ldots &
    \ldots & \ldots \\
    \hline
    C1 & 2 & 3 & 35 & 0 & 2 & $-4$ & \ldots & 1.30 & \ldots & \ldots &
    \ldots & \ldots \\
    C2 & 2 & 3 & 35 & 0 & 2 & $-3$ & \ldots & 1.30 & \ldots & \ldots &
    \ldots & \ldots \\
    C3 & 2 & 3 & 35 & 0 & 1.8 & $-2$ & \ldots & 1.30 & \ldots & \ldots &
    \ldots & \ldots \\
    C4 & 2 & 3 & 35 & 0 & 1.8 & $0$ & \ldots & 1.30 & \ldots & \ldots &
    \ldots & \ldots \\
    \hline
    D & \ldots & \ldots & \ldots & \ldots & 1.85 & \ldots & 1.5  &
    1.30 & \ldots 
    & \ldots & \ldots & \ldots \\
    E & \ldots & \ldots & \ldots & \ldots & 1.85 & \ldots & 1.0 & 1.30 & \ldots
    & \ldots & \ldots & \ldots \\
    F & \ldots & \ldots & \ldots & \ldots & 1.85 & \ldots & 0.5 & 1.30 & \ldots
    & \ldots & \ldots & \ldots \\
    G & \ldots & \ldots & \ldots & \ldots & 1.85 & \ldots & 0.25 &
    1.30 & \ldots
    & \ldots & \ldots & \ldots \\
    \hline
    W1 & 4 & 1 & 90 & 45 & 2 & \ldots & \ldots & 0.52 & 2.5 & 26.6 & 0.5 & \ldots \\
    SA & 4 & 1 & 90 & --45 & 2 & \ldots & \ldots & 0.33 & 4 & 26.6 & 0.5 & 25 \\
    SB & 4 & 1 & 90 & --45 & 2 & \ldots & \ldots & 0.33 & 4 & 63.4 & 0.5 & 25 \\
    
    \hline
\end{tabular}
  \label{tab:models}
\end{table*}

\subsection{Multilayer shock structure}
\label{sec:shock_structure}
\begin{figure*}
\includegraphics[width=120mm, keepaspectratio]{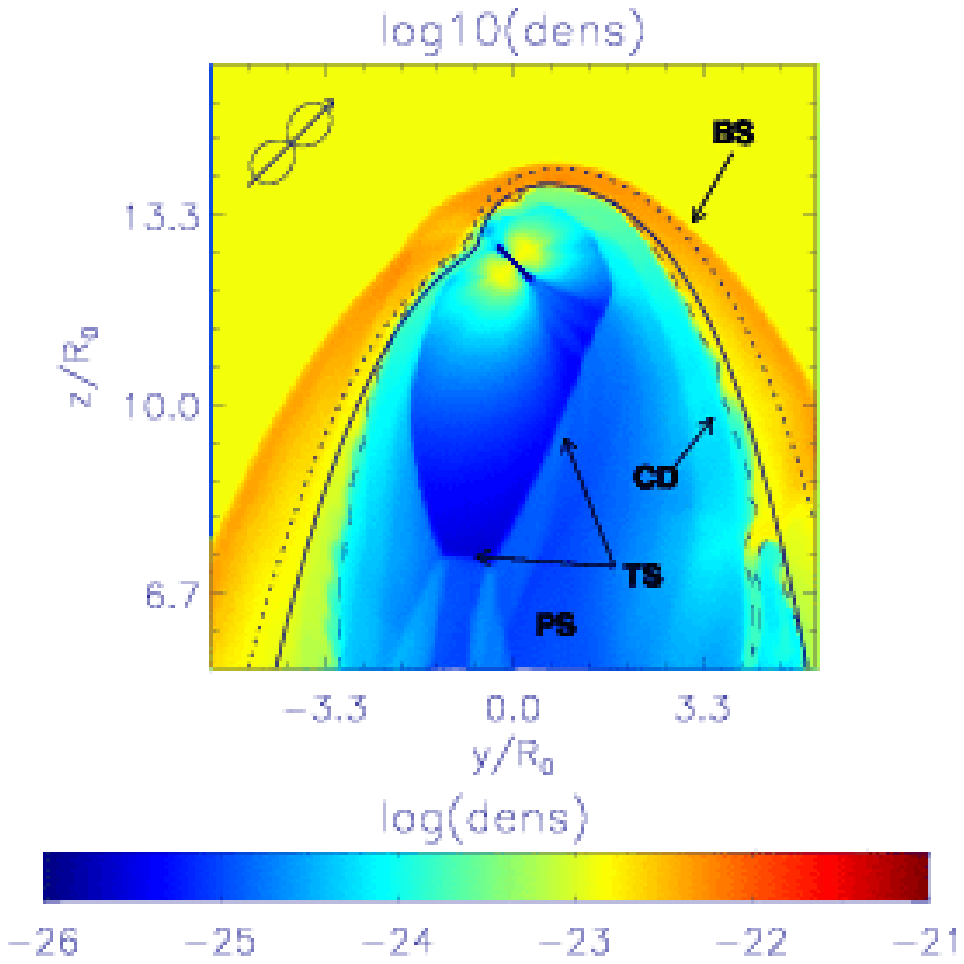}
\caption{$y$-$z$ section of model A. The colors denote the density in
  logarithmic scale, as indicated by the color bar. The characteristic
  multilayer structure, including the
  bow shock (BS), termination shock (TS), contact discontinuity (CD),
  and postshock region (PS), can be seen clearly. The solid curve shows
  the analytic approximation (\ref{eq:res:wilkin}), with $R_0$ scaled to match the
  CD (dashed curve), measured empirically by tracing the boundary
  between the two fluids in \textsc{Flash}. The dotted curve is the analytic solution
  scaled to match the BS. The projected symmetry axis and the wind momentum
  flux, drawn as a polar plot, appear in the top left corner of the figure.}
\label{fig:res:aniso_a_yz}
\end{figure*}

Figure \ref{fig:res:aniso_a_yz} shows the $y$-$z$ section of model
A, a typical anisotropic wind with $p=2$ and $c_2=3$ in which
$\mathbf{\Omega}$ is tilted by $45$\degr  with respect to
$\mathbf{v}_p$. This wind is pole-dominated, with zero momentum flux
at the equator, contrary to what is inferred from observation
\citep{Gaensler06a}. We postpone a detailed comparison between a pole-dominated
and an equatorially-dominated outflow to section \ref{sec:pole_vs_eq} and
discuss here only the shock structure, which is common to both cases.

All our simulations show the characteristic multilayer
structure in figure \ref{fig:res:aniso_a_yz}, described also by \citet{Swaluw03} and \citet{Bucc02}. The pulsar
wind inflates a wind cavity that is enclosed by a termination
shock (TS). The location of the TS behind the pulsar is determined by
the thermal pressure in the ISM \citep{Bucc02, Swaluw03}. Ahead of the pulsar, the termination
shock is confined by the ram pressure of the ISM.
The wind material is thermalised at the TS and fills a cylindrical
postshock region (PS), separated from the shocked ISM
by a contact discontinuity (CD).

In the case of a thin shock, the shape of the CD can be described 
analytically \citep[][see also appendix \ref{sec:app:analytic}]{Wilkin00}, with
a global length scale $R_0$ which is computed
by equating the ram pressures of the ISM and the wind:
\begin{equation}
  \label{eq:res:standoff}
  R_0 = \left(\frac{\dot{E}}{4 \pi \rho_m v_m^2 v_w}\right)^{1/2}.
\end{equation}
The analytic solution is depicted by a solid curve (scaled to match the CD) and by a dotted
curve (scaled to match the BS) in figure \ref{fig:res:aniso_a_yz}. As
discussed below (section \ref{sec:anisotropy:asymmetries}), it
describes the location of the CD and the BS, respectively, reasonably well.

For an isotropic wind, $R_0$ equals the standoff
distance, defined as the distance from the pulsar to the intersection
point of the pulsar's velocity
vector and the bow shock apex. For an anisotropic wind, however, the
standoff distance also depends on the exact distribution of momentum
flux and $R_0$ is merely an overall length-scale parameter \citep{Wilkin00}.

We can observe Kelvin-Helmholtz (KH) instabilites at the CD,
most prominently in the lower-right corner of figure \ref{fig:res:aniso_a_yz}. In the absence of
gravity, KH instabilities occur on all wavelengths, down to the grid
resolution, with growth rate merely proportional to wavelength
squared. Sure enough, the length-scale of the ripples in figure
\ref{fig:res:aniso_a_yz} is indeed comparable to
the grid resolution ($\Delta_x \approx 0.05 R_0$, $\Delta_y \approx
0.125 R_0$).

\begin{figure*}
  \includegraphics[width=168mm, keepaspectratio]{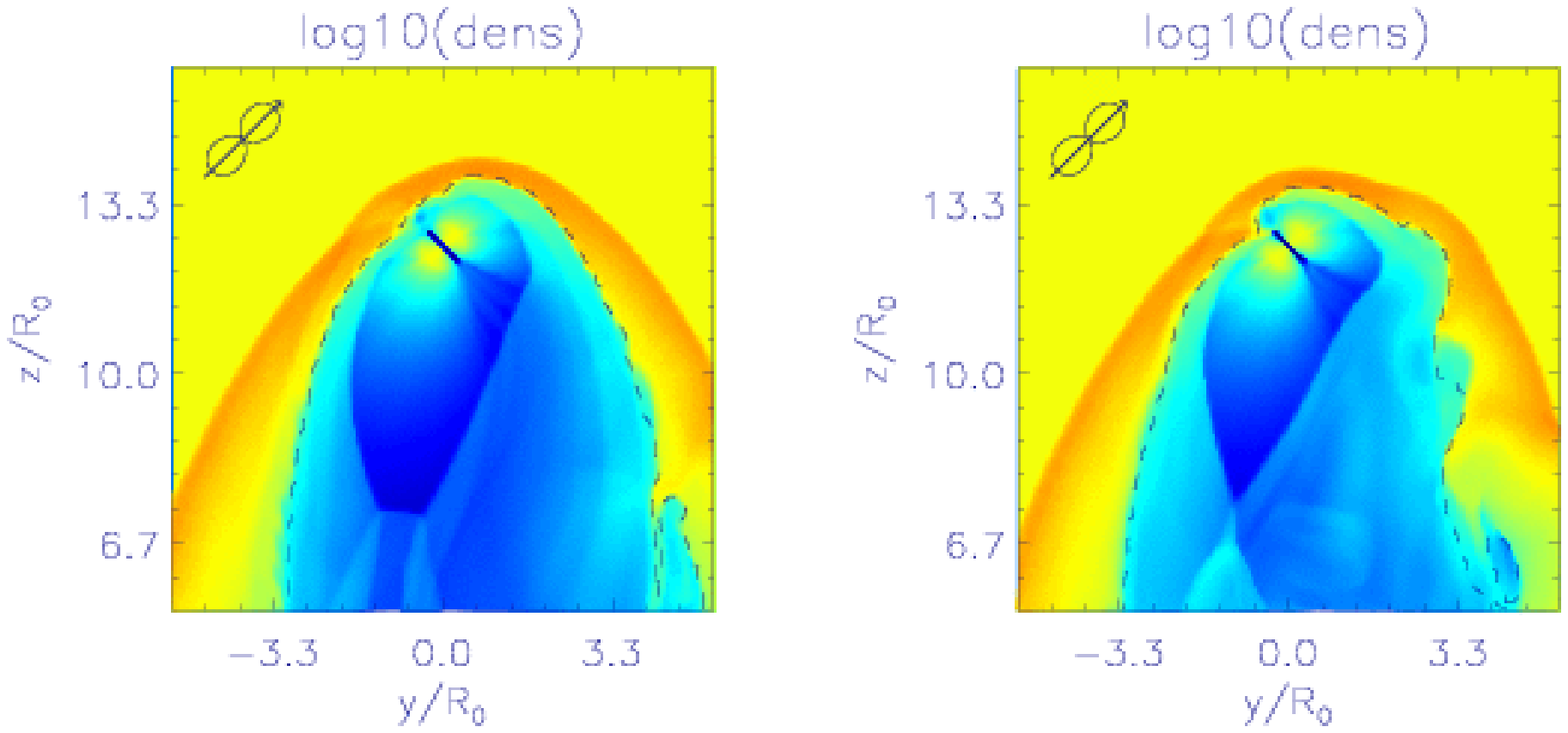}
  \caption{$y$-$z$ section of model A for $\gamma_1=5/3$ (left) and
    $\gamma_1=4/3$ (right), while $\gamma_2=5/3$. The shock morphology
  is the same. The system has not yet reached steady state, as
  can be seen by the ripples in the BS and CD.}
  \label{fig:setup:gammacomp}
\end{figure*}

Figure \ref{fig:setup:gammacomp} compares the results for model A with
$\gamma_1=5/3$ (left) and $\gamma_1=4/3$ (right), while $\gamma_2=5/3$
in both cases. As expected, the overall shock morphology
is weakly affected by adopting the adiabatic index $\gamma_1=4/3$ for
a relativistic gas, except for the ripples which arise because the simulation
has not yet reached a steady state.

\begin{figure}
  \includegraphics[width=84mm, keepaspectratio]{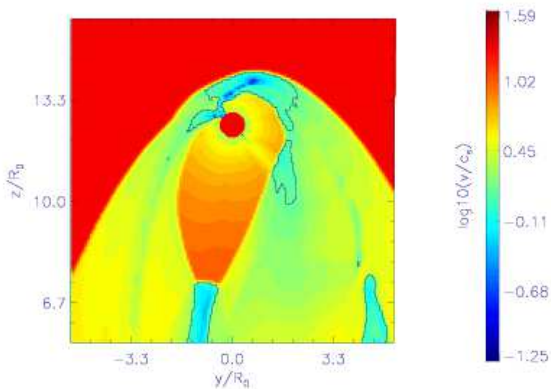}
  \caption{The ratio of total velocity $v=(v_x^2+v_y^2+v_z^2)^{1/2}$
    to adiabatic sound speed $c_s=(\gamma p/\rho)^{1/2}$ in a
    $y$-$z$ section of model A. The colors denote $v/c_s$ in logarithmic
    scale, as indicated by the color bar. The bulk flow is highly
    supersonic along, and nearly everywhere within, the
    boundaries. The solid lines denote the contour levels where $v/c_s=1$.}
  \label{fig:setup:soundspeed}
\end{figure}

\begin{figure*}
  \includegraphics[width=168mm, keepaspectratio]{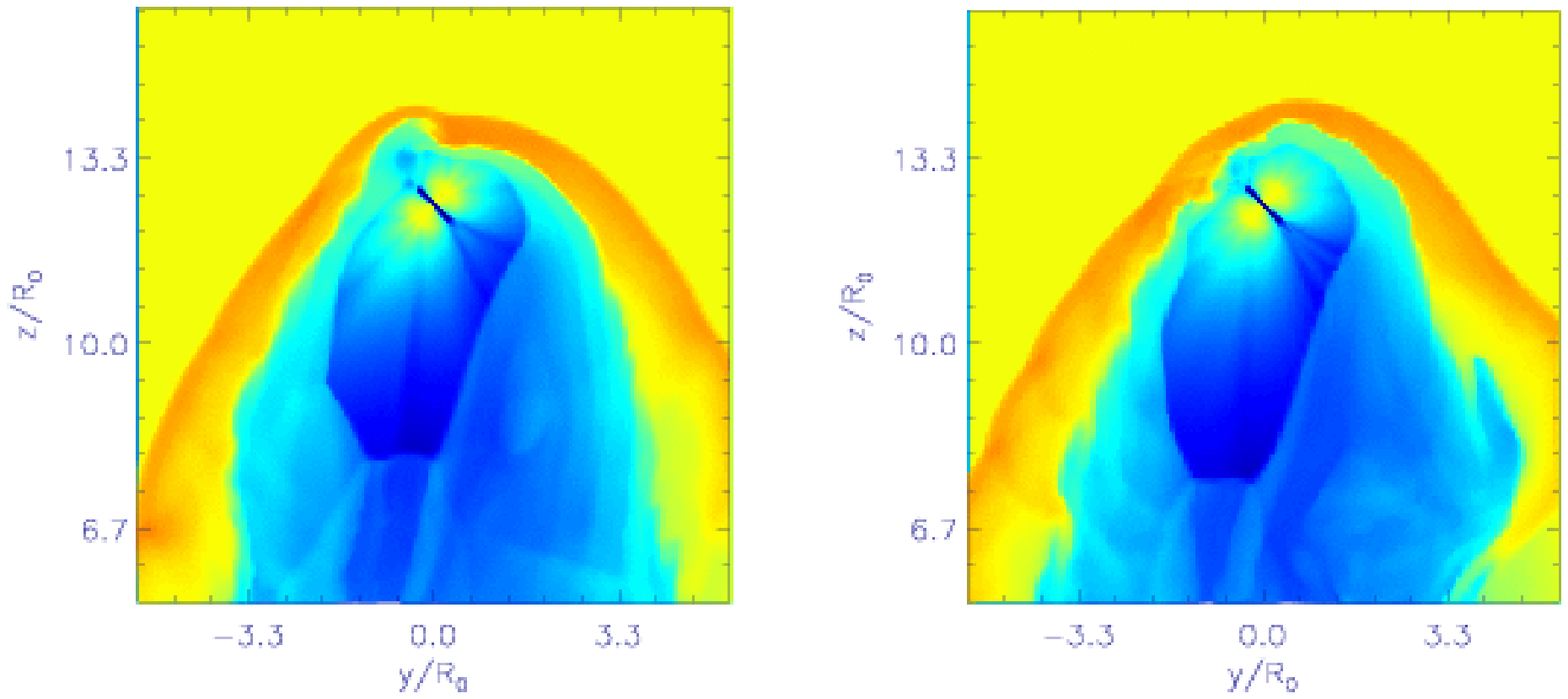}
  \caption{$y$-$z$ sections of model A with a computational domain of
    $x_\mathrm{max}-x_\mathrm{min}=y_\mathrm{max}-y_\mathrm{min}=z_\mathrm{max}-z_\mathrm{min}=10.6 R_0$
    (left) and the same model with a domain of
    $x_\mathrm{max}-x_\mathrm{min}=y_\mathrm{max}-y_\mathrm{min}=z_\mathrm{max}-z_\mathrm{min}=21.3 R_0$
    (right, zoomed-in to allow comparison with the left figure) at $t=3.38 \tau_0$ and $t=3.35 \tau_0$,
    respectively. The colors denote the density in 
    logarithmic scale as in figure \ref{fig:res:aniso_a_yz}. The shock
    morphology shows slight differences that can be attributed to the
    different evolution of the KH instabilities (due to the different
    resolution of the wind region and the time when the snapshot is taken).
    We do not observe any back reaction effects in the larger computational domain.}
  \label{fig:setup:smallhuge}
\end{figure*}

Figure \ref{fig:setup:soundspeed} displays the Mach number $v/c_s$
[where $c_s=(\gamma p/\rho)^{1/2}$ is the adiabatic sound speed] as a
function of position in a $y$-$z$ section of model A. The flow
is supersonic nearly everywhere along the boundaries and on the pulsar
side of the BS. The only exception is a few small regions of subsonic
flow in the PS (solid contours in figure \ref{fig:setup:soundspeed}) and we
do not expect the back reaction from these subsonic inclusions to affect
the flow structure. In order to verify this claim, we perform one simulation with the
parameters of model A and a larger computational domain
($x_\mathrm{max}-x_\mathrm{min}=y_\mathrm{max}-y_\mathrm{min}=z_\mathrm{max}-z_\mathrm{min}=21.3
R_0$, compared to
$x_\mathrm{max}-x_\mathrm{min}=y_\mathrm{max}-y_\mathrm{min}=z_\mathrm{max}-z_\mathrm{min}=10.6
R_0$ in figure \ref{fig:res:aniso_a_yz}) and compare the results in figure 
\ref{fig:setup:smallhuge}. We restrict the runtime to
$t=3.38 \tau_0$ (where $\tau_0$ is the time for the ISM to cross the
visible domain) to keep the computation tractable.
Although there are slight changes in the shock structure, which
can be attributed to the different resolution of the wind
region, we do not see any evidence for back reaction (e.g., matter or
sound waves travelling into the visible domain).

\subsection{Shock asymmetries}
\label{sec:anisotropy:asymmetries}

The analytic solution (\ref{eq:res:wilkin}) (solid curve in
figure \ref{fig:res:aniso_a_yz}) approximates
the shape of the CD (dashed curve) reasonably well. Here, and in all
the following plots, the CD in the simulation is defined by the condition
$X_2=0.9$ [see \eref{def_mass_fraction}]. Plainly, the CD in figure \ref{fig:res:aniso_a_yz} is
asymmetric. The kink that appears in the analytic solution at
the latitude where $\rho_w v_w^2$ is a minimum ($\theta=\pi/2$,
perpendicular to $\mathbf{\Omega}$ in the figure) is less distinct in
the simulations, because the thermal pressure in the wind and the ISM
smoothes out sharp features. \emph{Hence, even a highly
anisotropic wind does not lead to large, easily observable, kink-like
asymmetries in the CD}, an important (if slightly disappointing) point
to bear in mind when interpreting PWN observational data.

The point of the BS closest to the pulsar is no longer along
$\mathbf{v}_p$ but depends also on the latitude where the
momentum flux is a minimum. However, the BS itself is
also smoothed by thermal pressure so it is not a good gauge of the
underlying anisotropy. Equally, it is difficult to infer the
density of the ambient ISM from the location of the stagnation point, defined as
the intersection point of $\mathbf{v}_p$ with the BS, since
the orientation of $\mathbf{\Omega}$ cannot be inferred uniquely from
the shape of the BS.

The asymmetry of the BS in figure \ref{fig:res:aniso_a_yz}, which is more extended
towards the right side of the figure, is caused by
the excess momentum flux along $-\mathbf{v}_m$ on this side.
\citet{Bucc02} showed that the analytic formula \eeref{res:wilkin_simple}
accurately describes the CD for isotropic momentum flux. We
confirm here that the analogous formula \eeref{res:wilkin} accurately
describes the CD for
anisotropic momentum flux. But, as the wind ram
pressure decreases downstream from the apex, the thermal pressure
becomes increasingly important and the analytic solution breaks
down. Nevertheless, the asymmetry is preserved. Likewise, the
analytic solution for the BS (dotted curve)
deviates more strongly as one moves further downstream from the apex. Again, this
result is expected, since the analytic approximation relies on mechanical
momentum flux balance and breaks down as the thermal pressure
becomes increasingly important.

Momentum flux anisotropy affects the BS weakly, but it affects the
structure of the inner flow strongly. A
striking feature of figure \ref{fig:res:aniso_a_yz} is the shape of
the TS. It is highly asymmetric and elongated, roughly aligned with
$\mathbf{\Omega}$. As the rear surface of the TS lies well inside the
BS, its position is determined by the balance between the wind
ram pressure and the thermal pressure in the PS region, which is found
empirically to be roughly the same as in the ISM \citep{Bucc02}. This,
combined with the small momentum flux at $\theta=-\pi/2$ (running
diagonally from the upper left to bottom right in figure \ref{fig:res:aniso_a_yz}), explains
why the side surfaces of the TS are confined tightly, and hence its
characteristic convex shape.

\begin{figure}
\includegraphics[width=84mm, keepaspectratio]{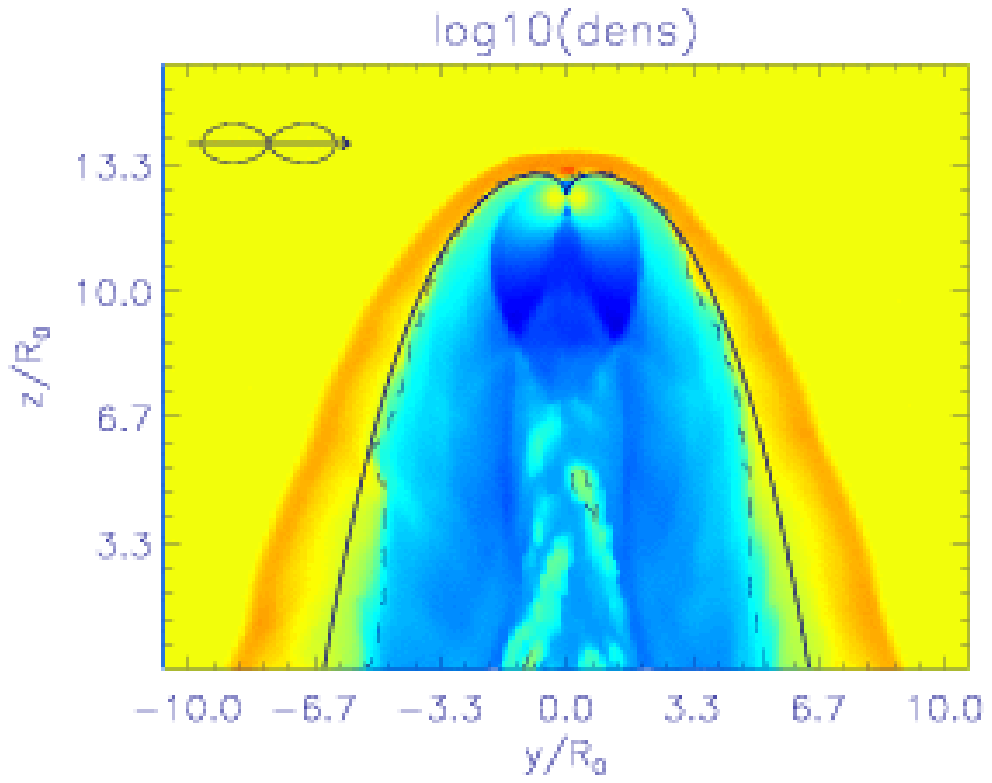}
\caption{$y$-$z$ section of model B. The colors denote the density in
  logarithmic scale, as defined in figure
  \ref{fig:res:aniso_a_yz}. The solid curve shows
  the theoretical solution (\ref{eq:res:wilkin}), scaled to match the
  CD (dashed curve). The solid and dashed curves overlap almost
  perfectly. The symbol in the top left indicates the projected
  wind symmetry axis and momentum flux (polar plot), as
  in the caption of figure \ref{fig:res:aniso_a_yz}. Note the
  butterfly-like TS and the BS almost touching the pulsar.}
\label{fig:res:aniso_b_yz}
\end{figure}

We now contrast the above results with a system that is axisymmetric around $z$.
Figure \ref{fig:res:aniso_b_yz} shows the $y$-$z$ section of model
B, which is the same as model A, except that $\mathbf{\Omega}$ is now
perpendicular to $\mathbf{v}_m$. Not surprisingly, the BS and CD are
axisymmetric. The analytic formula \eeref{res:wilkin} describes the CD
reasonably well, although the BS deviates slightly near the apex for the
pressure-related reasons discussed above. The butterfly-like shape of
the TS directly reflects the momentum flux and is a good observational
probe in this situation. Note that the CD and BS nearly touch the
pulsar at $\theta=-\pi/2$ (where the momentum flux is zero). This
illustrates that an estimate of $\rho_m$ based on
\eref{res:standoff} and the assumption of an isotropic wind
is too low.  Note again the KH instabilites at the CD and the blobs of
ISM matter in the PS region, which are numerical artifacts.

An example of the difficulties arising from the
degeneracy of the problem is PSR J2124$-$3358 \citep{Gaensler02}. The
PWN around this pulsar shows a clear asymmetry near the
apex. \citet{Gaensler02} showed that several combinations of an ISM
density gradient (perpendicular and parallel to $\mathbf{v}_m$), an
ISM bulk flow, and an anisotropic pulsar wind can explain the observed
anisotropy, but only in combination, not alone.

\begin{figure}
\includegraphics[width=84mm, keepaspectratio]{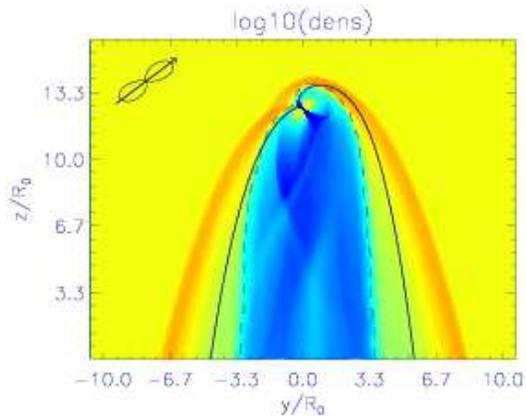}
\caption{$y$-$z$ section of model C. The colors denote the density in
  logarithmic scale, as defined in figure
  \ref{fig:res:aniso_a_yz}. The symbol in the top left indicates the
  projected wind symmetry axis and momentum flux (polar plot), as in the caption of figure
  \ref{fig:res:aniso_a_yz}. The solid curve shows
  the theoretical solution (\ref{eq:app:wilkinp4}), scaled to match the
  CD (dashed curve). The momentum
  flux in the wind is highly collimated, with $g_w \propto \cos^4 \theta$.}
\label{fig:res:aniso_c_yz}
\end{figure}

Figure \ref{fig:res:aniso_c_yz} demonstrates what the BS looks like
when the wind is highly collimated along $\mathbf{\Omega}$, i.e. a jet
($g_w \propto \cos^4 \theta$). Basically, the situation is similar to
model A (figure \ref{fig:res:aniso_a_yz}): the
analytic formula correctly predicts the shape of the CD. However, the kink at the
latitude of the momentum flux minimum is less visible than in figures
\ref{fig:res:aniso_a_yz} and \ref{fig:res:aniso_b_yz}. Moreover, the
wind cavity is smaller than in figure \ref{fig:res:aniso_a_yz}.
The TS approaches the pulsar more closely from behind, where there is less wind
momentum flux than in models A and B due to the high collimation.
The TS is also confined more tightly by the ISM ram pressure.

\subsection{Split-monopole versus wave-like wind}
\label{sec:pole_vs_eq}
In principle, the results above afford a means of distinguishing
between wind models with different momentum flux distributions.
In reality, however, the conclusion from section
\ref{sec:anisotropy:asymmetries} and figure \ref{fig:res:aniso_a_yz},
that the shapes of the CD and BS depend weakly on $g_w(\theta)$,
seriously hampers this sort of experiment. For example, the wave-like
dipole wind analyzed by \citet{Melatos96} and \citet{Melatos97} predicts
$g_w(\theta)=\cos^2 \theta+1$ (for a point dipole, as calculated in
appendix \ref{sec:app:momflux}), while the split-monopole wind
analyzed by \citet{Bogovalov99} predicts
$g_w(\theta)=\sin^2\theta+1/\sigma$, yet these models produce very
similar shapes for the CD and BS, as we now show.

\begin{figure}
\includegraphics[width=84mm, keepaspectratio]{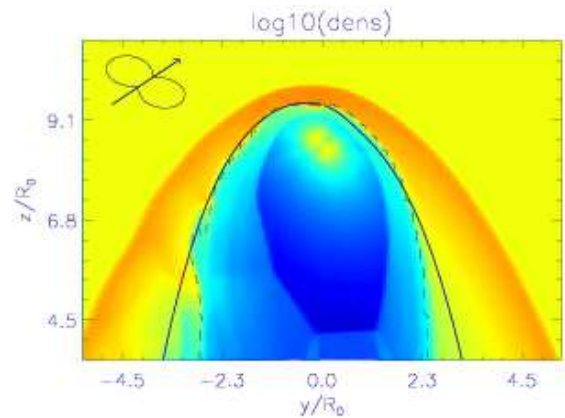}
\caption{$y$-$z$ section of model A1. The colors denote the density in
  logarithmic scale, as defined in figure
  \ref{fig:res:aniso_a_yz}. The symbol in the top left indicates the
  projected wind symmetry axis and the momentum flux (polar plot), as in the caption of figure
  \ref{fig:res:aniso_a_yz}. The solid curve shows
  the theoretical solution (\ref{eq:res:wilkin}), scaled to match the
  CD (dashed curve). This model
  describes a wind whose momentum flux is concentrated in the
  equatorial plane, with $g_w(\theta)=\sin^2\theta+7/3$}.
\label{fig:res:aniso_a1_yz}
\end{figure}

Consider an equatorially dominated wind, $c_2=-1$, implying
$g_w(\theta) = 7/3+\sin^2 \theta$.  This sort of wind is favored
by some PWN observations \citep{Chatt02}.
Figure \ref{fig:res:aniso_a1_yz} shows the $y$-$z$ section of
model A1. We can hardly see any
characteristic features in the bow shock. The CD is still described by the analytic solution. The
TS looks similar to the TS in model A, mirrored along the $x$-$z$
plane. Indeed, this simulation strongly resembles the situation of a
pole-dominated flux with $\lambda=135$\degr\ (cf. $\lambda=45\degr$ in
model A). We conclude that it will be hard to distinguish these cases observationally.

\begin{figure*}
\includegraphics[width=168mm, keepaspectratio]{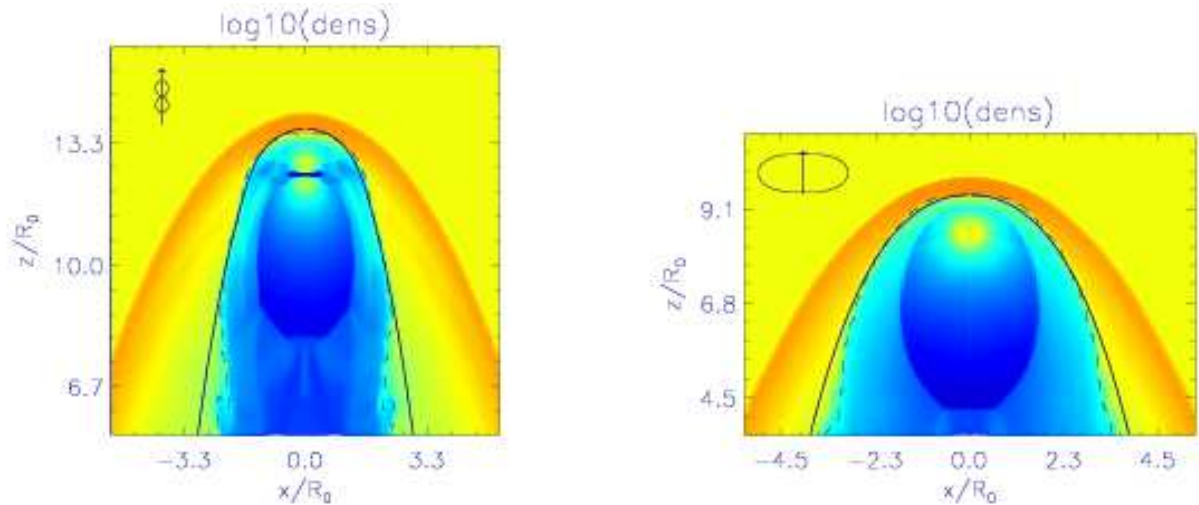}
\caption{$x$-$z$ section of model A (left) and model A1 (right). The colors denote the density in
  logarithmic scale, as defined in figure
  \ref{fig:res:aniso_a_yz}. The symbol in the top left indicates the
  wind symmetry axis, as in the caption of figure
  \ref{fig:res:aniso_a_yz}. The solid curve shows
  the theoretical solution (\ref{eq:res:wilkin}), scaled to match the
  CD (dashed curve).}
\label{fig:res:aniso_comp_yz}
\end{figure*}

Out of interest, a comparison between the $y$-$z$ section and $x$-$z$
section of these models can, in principle, reveal the underlying
wind structure. Such a comparison is depicted in figure
\ref{fig:res:aniso_comp_yz}. The pole-dominated wind (left panel)
has an anisotropic TS, while the TS in the equatorially-dominated
wind (right panel) resembles a completely isotropic
outflow. Unfortunately, such a comparison requires orthogonal lines of
sight and is not accessible by
observation. The computation of synchrotron emission maps
for both situations from different angles may provide additional
information, if Doppler boosting and column density effects break the
degeneracy in figure \ref{fig:res:aniso_a1_yz}. However, reliable
synchrotron maps require RMHD simulations and lie outside the scope of this paper.

\section{Cooling}
\label{sec:cooling}

\subsection{Numerical implementation}
The analytic solution derived by \citet{Wilkin96,Wilkin00} is exact in
the single-layer, thin-shock limit, where both the shocked ISM and the
wind cool radiatively (via spectral line emission and
synchrotron radiation respectively) faster than the
characteristic flow time-scale. In this section, we quantify how
rapidly such cooling must occur for the thin-shock approximation to be
valid. We also present some simulations of shocks where cooling is not
efficient and draw attention to how the structure of these shocks
differs from the predictions of \citet{Wilkin00}.

\textsc{Flash} implements optically thin cooling by adding a sink term
$-\Lambda(\mathbf{x},t)$ to the right-hand side of equation
(\ref{eq:setup:energy}). Radiation is emitted from an optically thin
plasma at a rate per unit volume $\Lambda(T)=X_0 (1-X_0) \rho^2
m_p^{-2} P(T)$, where $\rho$ is the density, $X_0$ is the ionisation fraction, $m_p$
is the proton mass, and $P(T)$ is a loss function that depends on
the temperature only  \citep{Raymond76, Cox69,Rosner78}. In our temperature
range, $10^4$ K $\leq T \leq 10^5$ K, hydrogen line cooling dominates and we can
write $P(T) \approx 10^{-22}$ erg s$^{-1}$ cm$^3$. The postshock
temperature $T_\mathrm{PS}$ can be extracted directly from the
simulation or estimated from the Bernoulli equation and entropy
conservation \citep{Bucc01}. With the latter approach, the result is
$k_\mathrm{B} T_\mathrm{PS}=3 m_p v_p^2/32$,
where $k_\mathrm{B}$ is Boltzmann's constant.

For a slowly moving pulsar, like PSR J2124$-$3358 which has
$v_p=61$ km s$^{-1}$ and $\rho_m=1.6 \times 10^{-24}$
g cm$^{-3}$ \citep{Gaensler02}, we find $T_\mathrm{PS}=4.2\times 10^4$ K and
$\Lambda=6.1 \times 10^{-21}$ erg s$^{-1}$ cm$^{-3}$, if we assume a moderate
ionisation fraction $X_0=0.2$. The internal energy per unit volume in the
postshock flow is $\epsilon = (\gamma-1)^{-1} (N_a  k_B/\bar{A}) T \rho =
1.4\times 10^{-11}$ erg cm$^{-3}$ where $\bar{A}=1$ g is the average
molecular mass of the hydrogen plasma, from which we deduce a cooling time
$\tau=\epsilon_p/\Lambda = 2.3 \times 10^9$ s. In
this object, the optical bow shock extends 65\arcsec or
$l=2.3 \times 10^{18}$ cm, implying a flow time $t_f=l/v_p=3.8 \times
10^{11}$ s and hence $\tau \ll t_f$. This example shows that cooling is
important not only in high-luminosity pulsars and in a high density
ISM \citep{Bucc01}, but also if the shock is sufficiently extended,
i.e. if the typical flow time, depending on $\mathbf{v}_p$, approaches
the typical cooling time, which is a function of $\mathbf{v}_p$ and $\rho_m$.
While \citet{Bucc01} argued that the typical cooling length scale is
the width of the BS plus the distance between BS and CD, we include
the whole visible BS, allowing a longer time for the shocked ISM to
cool. This is justified because the material in the post-BS region
continues to cool and participate in the dynamics of the system.

We can perform a similar calculation for a fast pulsar, like
PSR J1747$-$2958 (the Mouse) which has $v_p=600$ km
s$^{-1}$ and $\rho_m=5.0 \times 10^{-25}$ g cm$^{-3}$
\citep{Gaensler02}, corresponding to $T_\mathrm{PS}=4.1\times 10^6$ K and
$\Lambda=1.7 \times 10^{-24}$ erg s$^{-1}$ cm$^{-3}$. The cooling
efficiency here is lower than in the previous example, since hydrogen
line cooling is slower at high temperatures
\citep{Raymond76} and $\rho_m$ is lower. The cooling time $\tau=1.5
\times 10^{14}$ s is long compared to the flow time-scale $t_f=1.5\times
10^{10}$ s; the length of the BS is $l=0.3$ pc.

To treat H$\alpha$ cooling properly, one must solve the Boltzmann equation
for the neutral atom distribution as a function of position, including
the ion-neutral reactions, a calculation which lies beyond the
scope of this article. Instead, in order to quantify the overall effects of
cooling and verify the validity of the analytic thin-shock approximation,
we parameterize the cooling function as
\begin{equation}
  \label{eq:setup:cooling}
  \Lambda(\mathbf{x},t)= \Lambda_0 f \rho \epsilon(\mathbf{x},t),
\end{equation}
with $\Lambda_0=10^{-7}$ s$^{-1}$, such that the dimensionless
parameter $f$ controls the local cooling time through $\tau = (\Lambda_0 f)^{-1}$
(which is then spatially uniform by construction, a crude approximation).

\subsection{Shock widths}

\begin{figure*}
\includegraphics[width=168mm, keepaspectratio]{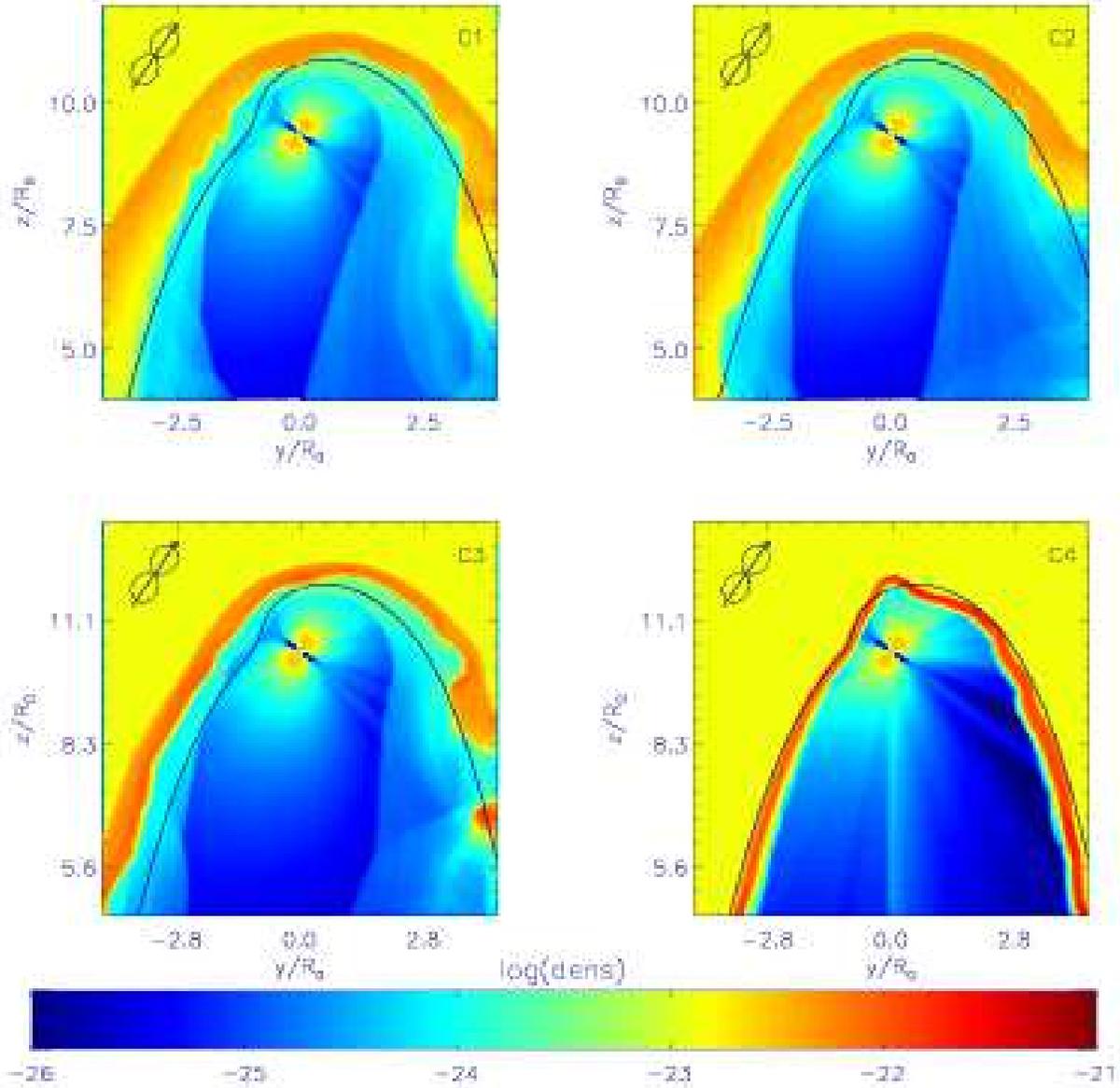}
\caption{$y$-$z$ section of models C1 ($f=10^{-4}$, upper left), C2
  ($f=10^{-3}$, upper right), C3 ($f=10^{-2}$, bottom left), and C4 ($f=1$,
  bottom right). The colors denote the density in
  logarithmic scale as indicated by the color bar (g cm$^{-3}$), while
  the solid curve shows the analytic solution (\ref{eq:res:wilkin}).
  The symbol in the top left indicates the projected wind symmetry
  axis and momentum flux (polar plot), as
  in the caption of figure \ref{fig:res:aniso_a_yz}. For more
  efficient cooling, e.g. C4, the TS lengthens
  and the BS becomes thinner, until the TS and BS are no longer
  separated. The \citet{Wilkin00} solution (solid curve) matches the bow shock
  nearly exactly (bottom right), with a deviation of $<8$\% from the
  actual BS position. Here, the bumps at the BS are an artifact
  of the discrete grid. They are not seen in C1--C3, where
  small density fluctuations are smoothed by the thermal
  pressure. However, C1--C3 exhibit strong KH instabilites at the BS.}
\label{fig:cool:coolseries}
\end{figure*}

Figure \ref{fig:cool:coolseries} shows the $y$-$z$ sections of models
C1 to C4. These four
models are identical ($g_w \propto \cos^2 \theta$, $\lambda=35$\degr;
see table \ref{tab:models}), except that $f$
varies from $10^{-4}$ (C1) to $10^0$ (C4). For more efficient cooling,
the BS becomes thinner (compare upper-left and bottom-right panels in
the figure). Similarly, the TS becomes more extended with increasing
cooling efficiency, until the wind cavity
fills up the BS interior and the TS becomes indistinguishable from the
BS for highly efficient cooling (bottom-right panel).
Since the pressure in the PS region approximately
equals the thermal ISM pressure, and since efficient cooling reduces
the PS pressure, the wind ram pressure drives the termination shock
outward as $f$ increases, as seen in figure
\ref{fig:cool:coolseries}. On the other hand, the thickness of the
bow shock is controlled by the ratio between the thermal and ram
pressures in the ISM. The BS region therefore becomes thinner for
more efficient cooling, as we pass from model C1 to C4. For $f=1$, the TS and BS
almost touch and the  one-layer, thin-shock analytic solution
given by equation (\ref{eq:res:wilkin}) applies as in the lower-right panel of
figure \ref{fig:cool:coolseries}.

\section{Smooth ISM Density Gradient}
\label{sec:densgrad}
Density gradients in the ISM have been invoked to explain the
asymmetric features observed in the bow shocks
around PSR B0740$-$28, PSR B2224+65 (the Guitar nebula), and PSR J2124$-$3358 \citep{Jones02, Chatt02,
  Gaensler02}. The PWN around PSR B0740$-$28 is shaped like a key hole, with a
nearly circular head and a divergent tail, starting $\sim 20 R_0$ downstream
from the apex. The Guitar nebula consists of a bright head, an
elongated neck, and a limb-brightened body. The PWN around PSR J2124$-$3358
exhibits an asymmetric head and a distinct kink $\sim 10 R_0$ downstream
from the apex.

\subsection{Numerical implementation}
To explore the effect of a smooth density gradient, we carry out a set of four simulations (D--G in
table \ref{tab:models}) in which the ISM has an exponential density profile
\begin{equation}
  \rho_m(y)=\rho_{m,0}\exp(-y/H),
\end{equation}
where $H$ is a length scale, and $\rho_{m,0}$ substitutes for
$\rho_m$ in table \ref{tab:models}. We generally choose $H \sim R_0
\sim 0.01$ pc, consistent with the wavelength of turbulent fluctuations in
the warm ISM \citep{Deshpande00}. The gradient is perpendicular to the
pulsar's motion. The density profile is initialized
and then maintained at later times by setting $\rho_m(y)$ at
the inflow boundary at $z=z_\mathrm{max}$
to $\rho_m(y)=\rho_{m,0} \exp(-y/R_0)$, and choosing $p_m$ and $\epsilon_m$ consistently.
The time-scale over which the associated pressure gradient smoothes
out $\rho$, $H/c_m \sim 10^{10}$ s, exceeds the
flow time-scale, $t_f = (z_\mathrm{max}-z_\mathrm{min})/v_m \sim
6\times 10^{9}$ s, so the density gradient remains nearly constant.

\begin{figure*}
\includegraphics[width=168mm, keepaspectratio]{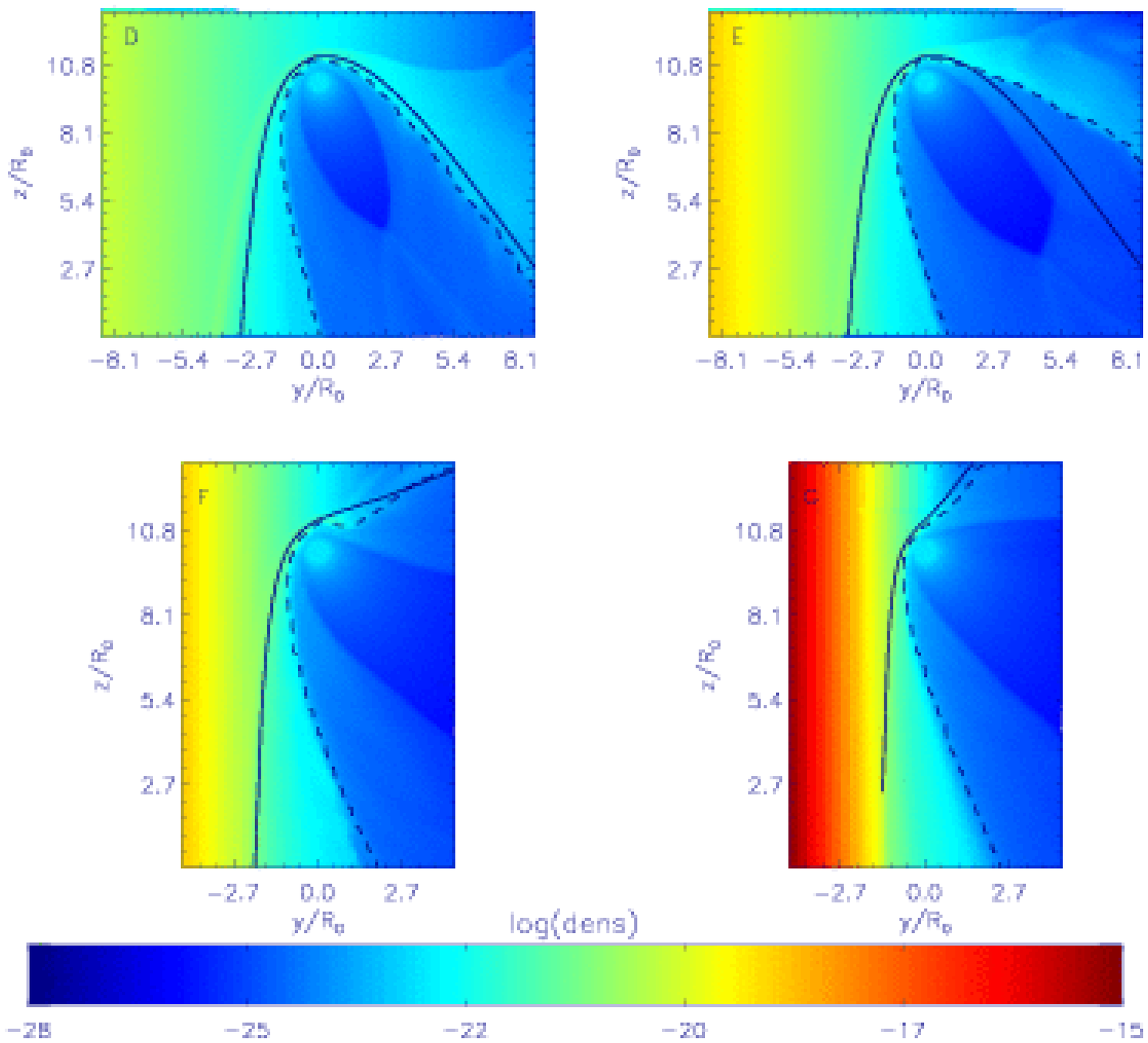}
\caption{$y$-$z$ sections of model D ($H=1.5 R_0$, upper left), E
  ($H=1.0 R_0$, upper right), F ($H=0.5 R_0$, lower left), and G
  ($H=0.25 R_0$, lower right). The colors denote the density in
  $\log_{10}$(g cm$^{-3}$); note that the color scale differs from that in the
  other figures. The solid curve shows
  the theoretical solution (\ref{eq:res:wilkindens}), scaled to match the
  CD (dashed curve). The analytic approximation clearly breaks down
  for models D--G (see text). The BS deviates from the CD where the ISM ram
  pressure exceeds the thermal pressure. Axes are labelled in
units of $R_0=1.85\times10^{16}$ cm.}
\label{fig:res:densgrad_series_yz}
\end{figure*}

The analytic solution for the shape of a thin bow shock in an
exponential density gradient is given by \eref{res:wilkindens}.

\subsection{Tilted shock}
\label{sec:densgrad:tilt}
Figure \ref{fig:res:densgrad_series_yz} shows the $y$-$z$ sections of
models D--G, together with the analytic solution for the BS (solid curve) and the
CD (dashed curve). In these models, the exponential density profile
changes from shallow ($H=1.5 R_0$ for model D, upper-left panel) to steep
($H=0.25 R_0$ for model G, bottom-right panel). One distinctive feature
is that the low-density side ($y>0$) of the shock
structure, where the CD, TS, and PS are situated, is tilted with respect
to $\mathbf{v}_m$ by an angle of up to 180\degr\ for $H/R_0=0.25$.
This occurs because the ISM ram pressure, $\rho_m v_m^2$, is
dominated by the wind ram pressure for $y>0$, pushing the
CD further away from the pulsar. In fact, this part of the CD is
approximated well by the analytic formula
(\ref{eq:res:wilkindens}) \citep{Wilkin00}. The BS separates from the CD and broadens as the
thermal pressure of the ISM in the low-density region 
increases relative to the ISM ram pressure.

On the high-density side ($y<0$) of the
bow shock, the opposite situation prevails. Here, the ram pressure of the
ISM exceeds its thermal pressure, so
that the BS is well approximated by the thin-shell solution. The
CD curves towards the lower density region, pushed by
the thermal ISM pressure. The opening angle of the CD increases
as $H$ decreases: we find 40\degr, 60\degr, 103\degr, and 124\degr\ for
models D, E, F, and G respectively, as the ISM ram pressure (for $y>0$)
decreases with increasing $y$.

Applying the above results to observations of PWN, we expect the
H$\alpha$ surface brightness ($\propto$ column density) to
mimic the volume density contours as in Figure \ref{fig:res:densgrad_series_yz}, especially
where the thickness of the BS increases from $y<0$ to $y>0$. Such a
variation in the H$\alpha$ flux has not been observed in the PWN
around PSR B0740$-$28, PSR B2224+65, and PSR J2124$-$3358 \citep{Jones02, Chatt02,
Gaensler02}. It therefore seems unlikely that a smooth
density gradient is responsible for the peculiar morphologies observed
in these objects \citep{Chatterjee06}.

\section{Wall in the ISM}
\label{sec:wall}

The ISM is inhomogenous on length scales from kpc down to AU
\citep{Deshpande00}. Hence a fast pulsar is likely to encounter
singular obstacles or barriers, such as the edges of O-star
bubbles \citep{Naze02}, which effectively act as ``walls''. A
wall was invoked to explain the kink observed in the nebula around PSR J2124$-$3358
\citep{Chatterjee06}. Unlike when a pulsar interacts with
its supernova remnant \citep{Swaluw03}, the pulsar does not always
strike the wall head on, giving rise to a
highly asymmetric interaction.

\subsection{Numerical implementation}
In order to model the wall, we introduce the (signed) coordinate
$\varpi$ along the direction defined by $\cos \alpha \, \hat{\mathbf{y}}
- \sin \alpha \, \hat{\mathbf{z}}$, perpendicular to the wall
plane. In model W1, the wall is infinitely extended (i.e., a
ramp). Its profile can be written as
\begin{equation}
  \label{eq:anisotropy:infwall}
  \rho(\varpi) = (\rho_\mathrm{wall}-\rho_m)[1+\exp(\varpi/\Delta_r)]^{-1}
  +  \rho_m.
\end{equation}
In models S1 and S2, we truncate the wall on one side. Its profile can
be written as
\begin{eqnarray}
  \label{eq:anisotropy:finitewall}
  \rho(\varpi) & = & \rho_m +
  (\rho_\mathrm{wall}-\rho_m)[1+\exp(\varpi/\Delta_r)]^{-1} \nonumber \\
  & & + \theta(-\varpi-8\Delta_r) (\rho_\mathrm{wall}-\rho_m) \exp[(\varpi+8
    \Delta_r)/\Delta_d], \nonumber \\
\end{eqnarray}
where $\theta(\varpi)$ is the usual Heaviside function, defined as
$\theta(\varpi)=\{0 (\varpi<0), 1 (\varpi\ge0)\}$. In this
notation, $\varpi = 0$ denotes the center of the smooth rise in
equations (\ref{eq:anisotropy:infwall}) and (\ref{eq:anisotropy:finitewall}).
The maximum density is $\rho_\mathrm{wall}=1.30 \times 10^{-23}$ g cm$^{-3}$.

\begin{figure*}
\includegraphics[width=168mm, keepaspectratio]{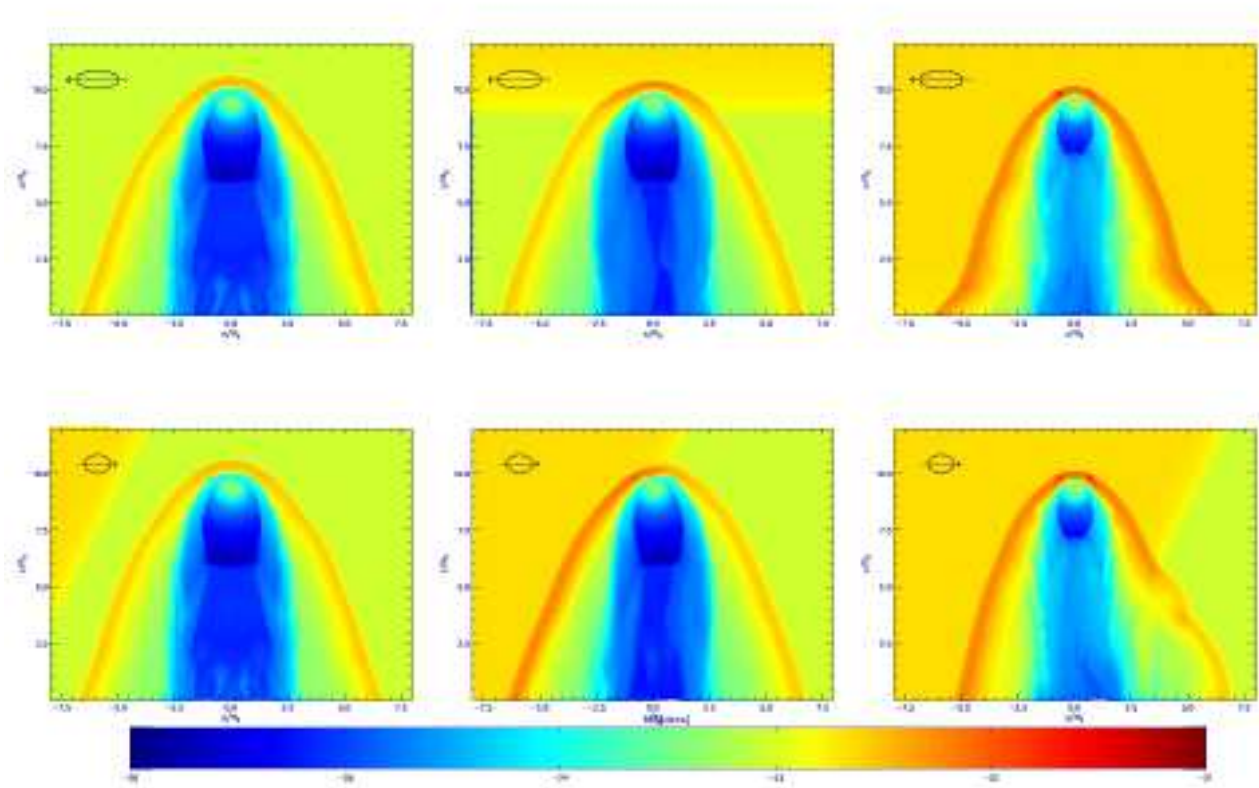}
\caption{Time series of $x$-$z$ (top row) and $y$-$z$ (bottom row)
  sections of model W1. The columns are at times $t=1.7 t_f$ (left),
  $t=2.5 t_f$ (middle), and $t=3.4 t_f$ (right), where $t_f=6\times
  10^9$ s is the time for the ISM to cross the simulation volume. The
  colors denote the density in $\log_{10}$ (g cm$^{-3}$), while the symbol in
  the top left indicates the projected wind symmetry axis and momentum flux (polar plot), as
  in the caption of figure \ref{fig:res:aniso_a_yz}. The density of the
  wall is 4 times higher than the ISM density.   The
  momentum flux is anisotropic ($\lambda=90$\degr) and $\mathbf{\Omega}$
  is tilted towards the observer ($\phi=45$\degr). The bow shock
  is not disrupted by the wall and shows characteristic kinks and
  Kelvin-Helmholtz instabilities.}
\label{fig:res:denswallseries}
\end{figure*}

\subsection{Kinks in the bow shock}
In this section, we examine the effect of a wall on the shock
structure and work out how PWN observations probe inhomogenities in
the ISM. We consider both a ramp-like [equation
  \eeref{anisotropy:infwall}] and a truncated [equation
  \eeref{anisotropy:finitewall}] wall.

Figure \ref{fig:res:denswallseries} shows
a simulation (model W1) where the pulsar hits a ramp-like wall with
$\alpha=26$\degr\ and density contrast
$\eta=\rho_\mathrm{wall}/\rho_m=2.5$. The wind is asymmetric, with
$\lambda=90$\degr and $\mathbf{\Omega}$ tilted towards the observer
($\phi=45$\degr), resulting in a slightly asymmetric PS flow. The top
and bottom rows of figure \ref{fig:res:denswallseries} depict
snapshots of $x$-$z$ and $y$-$z$ sections respectively. The $y$-$z$
sections demonstrate the asymmetry best. On impact, the BS is
compressed by the increased ram pressure of the wall. A kink appears at the
transition point between the wall and the ambient ISM. Both the
gradient in H$\alpha$ brightness and the kink should be observable in principle. As the pulsar proceeds into the
high-density region (see middle panels of Figure \ref{fig:res:denswallseries}), the CD approaches the pulsar
due to the increased ISM ram pressure. Likewise, the ISM thermal
pressure increases with increasing $\rho_m$ and, by the argument given
in section \ref{sec:anisotropy:asymmetries}, the rear surface of the
TS comes closer to the pulsar. Just as for a smooth ISM density gradient (cf. Figure
\ref{fig:res:densgrad_series_yz}), the shocks are tilted towards the
low density side. When the pulsar proceeds into the densest part of the ISM,
the BS, CD, and TS approach even closer to the pulsar until they finally
reach the state in the rightmost panels.

In this stationary state, KH instabilites triggered by
perturbations during the collision with the wall occur at the right hand
surfaces of the CD. In the absence of gravity, the KH instability occurs
for all wave numbers and hence on all length-scales down to the grid
resolution ($\sim 0.1 R_0$ in this simulation). However, the collision
with the wall introduces a velocity perturbation that is large ($\sim v_m$) compared to the
numerical noise. This explains why we see the
instability in the right panels but not the left panels, where the
wall has not yet reached the BS. Indeed, the length scale of
the density rise due to the wall agrees with the wavelength of the observed KH
instability ($\sim R_0$).

In the lower middle panel, the CD downstream from the apex
is slightly lopsided towards the higher density (left) side. This
occurs because the PS flow from the side surfaces of the TS travels faster than
the ISM. The region inside the CD therefore loses pressure
support by the shocked wind material, and the shocked ISM material
pushes into the space. The same effect can be seen in the
lower right panel on the lower density (right) side, where the KH
instabilities favor the right-hand side. The top row ($x$-$z$
sections) basically resembles a head-on collision, similar to the
interaction of a PWN with the pulsar's supernova remnant
\citep{Swaluw03}. Note also the
broadening of the lower part of the BS (top right panel), which
demonstrates the delay ($\sim 5 R_0/v_m$) before the shock adjusts to
the higher external density $\rho_m$.

The H$\alpha$ emission emanates chiefly from the region between the
BS and the CD \citep{Bucc01}, which is only moderately affected by the
passing wall. Furthermore, ISM inhomogenities often consist of highly ionised
matter \citep{Chatterjee06}, so that the H$\alpha$
luminosity observed as the wall passes is dominated by the afterglow of
the neutral part of the non-wall ISM. Negligible brightness
variations are observable in either case. Nevertheless, as we discuss in section
\ref{sec:discussion}, a comprehensive treatment of the emission requires the
solution of the Boltzmann equation for the neutral and ionized
species, a task beyond the scope of this article.

\begin{figure*}
\includegraphics[width=168mm, keepaspectratio]{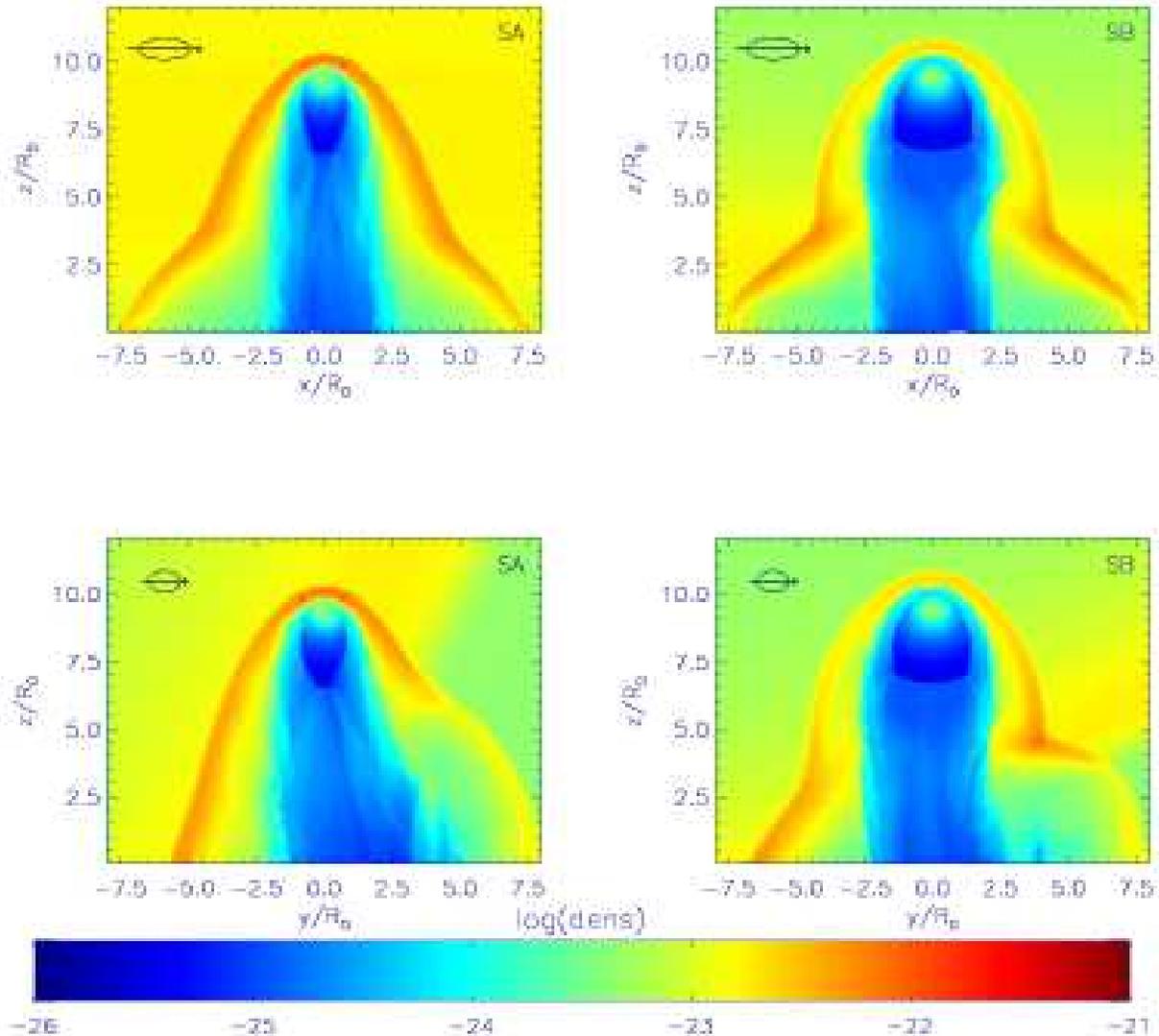}
\caption{$x$-$z$ (top row) and $y$-$z$ (bottom row) sections of model
  SA (left column) and model SB (right column). The colors denote the density in $\log_{10}$ (g cm$^{-3}$), while
  the symbol in   the top left indicates the projected wind symmetry axis and momentum flux (polar plot), as
  in the caption of figure \ref{fig:res:aniso_a_yz}. In these
  simulations, the pulsar travels into a wall with a finite width. The
  momentum flux is anisotropic ($\lambda=90$\degr) and $\mathbf{\Omega}$
  is tilted away from the observer ($\phi=-45$\degr).}
\label{fig:res:denswallsaandsb}
\end{figure*}

In models SA ($\alpha=26.6$\degr) and SB ($\alpha=63.4$\degr),
displayed in figure \ref{fig:res:denswallsaandsb}, the
density ramp is replaced by a wall of finite width, which is truncated as in
\eref{anisotropy:finitewall}. The top and
bottom row again shows $x$-$z$ and $y$-$z$ sections respectively. As
above, the $x$-$z$ sections resemble a head-on collision. The base of
the BS broadens in the low-density region and narrows at
the density maximum. Above the wall, the BS readjusts to the lower ISM
density. Due to the smooth gradient ($\Delta_d \gg \Delta_r$), the BS
looks ``egg-shaped'', an effect that can been seen more clearly in the
upper-right panel, where the inclination of the wall is lower
($\alpha=64$\degr). The kink is also more distinct,
as seen in the $y$-$z$ sections (lower panels), because the BS expands
in the low density region above the wall. The CD and TS approach the
pulsar for the same reasons discussed above, but the smooth
gradient is responsible for the bulbous shape. Again, as in figure
\ref{fig:res:denswallseries}, we can see in
the lower-right corners of the lower panels how the shocked wind
material inside the CD loses pressure support and the shocked ISM
material diffuses in. 

\section{Discussion}
\label{sec:discussion}
In this paper, we perform three-dimensional simulations of pulsar bow shocks
in the ISM under a variety of asymmetry-inducing conditions: an
anisotropic pulsar wind, an ISM density gradient, and an ISM wall. We
validate the analytic solution of \citet{Wilkin96, Wilkin00} for an anisotropic wind
and show that it remains a reasonable approximation for the thick-shock
case, with a deviation of $\sim 10$\% within $\sim 5 R_0$ of the
apex. We show that the basic multilayer shock structure (TS, CD and BS)
obtained by \citet{Bucc02} and \citet{Bucc05} for an isotropic
wind is preserved in an anisotropic wind as well. We find that the
thermal pressure smoothes the BS. The underlying anisotropy of the
momentum flux thus remains concealed, making it difficult to gauge
the ISM density from the stagnation point, since generally neither the
distribution of the momentum flux nor the orientation of the pulsar
spin axis is known.

Contrary to previous estimates \citep{Bucc01}, we show that cooling
can be important for slow pulsars in a low-density ISM, such as PSR
J2124$-$3358. This is because the shocked ISM continues to take part in the
dynamics of the system and thus influences the overall shape. There are observational
consequences: cooling tends to increase the overall size of the
TS while the BS becomes thinner. Naively, a thinner BS is expected to be
brighter due to limb brightening. However, in order to quantify the effects of
cooling on the optical emission, it is necessary to solve the
collisionless Boltzmann equation for the neutral hydrogen atoms, a
task beyond the scope of this paper.

Our simulations are non-relativistic and do not include the
effects of a magnetic field. \citet{Bucc05} showed that a proper
relativistic treatment increases the KH instabilities arising at the
CD by enhancing the velocity shear at the
interface. In principle, therefore, shocked ISM material can contaminate
the wind material, altering the inner flow structure and
perhaps modulating the synchrotron flux in time, although no simulation
has demonstrated the latter effect to date. A magnetic field affects the flow
more drastically, in two ways. First, the standoff distance increases because the magnetic
field adds to the pressure at the leading surface. This is
a runaway process in the simulations of \citet{Bucc05}, who
neglected resistivity. Second, the TS becomes convex because the
magnetic pressure is not uniform. The azimuthal magnetic field
builds up near the pulsar, increasing the magnetic
pressure. This effect is strongest at $z_p$ and decreases as $z$
decreases.

An interesting question is whether the magnetic field in
PWN can produce the torus-ring structure found by \citet{Komissarov03,Komissarov04} and
\citet{DelZanna04}, where a jet is formed by circulating back flow of the
shocked wind material colliminated subsequently by the magnetic
field. This circulation takes place outside the TS and is suppressed by strong
confinement in PWN. However, a detailed understanding
of the magnetic field structure when the momentum flux is asymmetric
must await future simulations.

PWN are commonly detected as radio or X-ray synchrotron sources. Charged wind leptons
are accelerated at the termination shock and subsequently gyrate in
the local magnetic field, emitting synchrotron radiation. Typically,
the emission is observed from three different zones:
the TS itself, the PS region, and the head of the BS. For an
identification of these zones in the Mouse, see
\citet{Gaensler04}. A further theoretical study of the synchrotron
emission will be undertaken in a forthcoming paper
\citep{Chatterjee06}. Here, we simply make a few qualitative points.

\begin{enumerate}\item Generally, we observe a variety of TS shapes from
inclined tori (Crab Nebula, where the emission is equatorially concentrated)
to cylinders (as seen in the Mouse, for a nearly
isotropic wind). The emission from the TS depends
strongly on the anisotropy of the wind.  The more peculiar the shape, the more
information it can provide about the inclination of the spin axis with respect to
the direction of motion as well as the angular distribution of wind
momentum flux. The asymmetry in the TS (figures \ref{fig:res:aniso_a_yz}
and \ref{fig:res:aniso_c_yz}) directly corresponds to the axis of
maximum emission.
\item For isotropic wind emission, one expects the PS flow to be observed as
an uncollimated X-ray tail \citep{Bucc02}. However, if the wind is anisotropic
(cf. Figures \ref{fig:res:aniso_a_yz}--\ref{fig:res:aniso_c_yz}), our
simulations predict overluminous and underluminous regions (corresponding to
high and low densities) in the PS flow, even two separated tails in
extreme cases (as in figure \ref{fig:res:aniso_b_yz}).
\item The observed appearance of the BS is strongly influenced by the
emission from particles accelerated at the head of the BS and
subsequently swept back along the CD. Since only the flow component normal to the shock
slows down, the flow in the swept-back region is generally trans-relativistic
($\sim c$). If this part of the PS flow
[region B1 in \citet{Chatterjee06}] is laminar along the CD, Doppler
beaming may render the emitted radiation invisible \citep{Bucc05}. An anisotropic wind,
however, may have a sufficiently large velocity component towards the
line of sight to allow detection \citep{Chatterjee06}. In the
case of a jet-like wind, the BS can exhibit a one-sided X-ray tail
whose receding half is suppressed by Doppler beaming,
as in the PWN around PSR J2124$-$3358 \citep[][whose model B corresponds
to our model SA]{Chatterjee06}. 
\end{enumerate}

Some PWN are also detected in neutral hydrogen emission lines
\citep{Gaensler06a}. Neutral hydrogen atoms are
excited collisionally or by charge exchange and then de-excite radiatively in the region between the BS
and CD. Hence, the H$\alpha$ luminosity depends crucially on $\mathbf{v}_p$ and
$X_0$. A proper kinetic (Boltzmann) treatment
incorporating ion-neutral reactions is discussed by \citet{Bucc01}.
It is clear that purely hydrodynamical simulations like ours cannot predict
the observed surface brightness reliably. However, in a paper in
preparation \citep{Chatterjee06}, we show that the overall shape of the
bow shock is predicted reliably.

If high-resolution observations of the shock apex can resolve the
characteristic kink in the BS at the latitude where the momentum flux
is a minimum, they will shed light on the angular distribution of the pulsar wind momentum
flux and hence the electrodynamics of the pulsar magnetosphere
\citep{Bogovalov99, Komissarov03,Komissarov04, Melatos96, Melatos97,
Melatos02}. However, it is generally difficult to distinguish between
equatorially dominated (e.g., split-monopole) and pole-dominated
(e.g., wave-like dipole) wind models, beacuse the shock structure is
degenerate with respect to several parameters and does not depend
sensitively on the momentum flux distribution (see sections
\ref{sec:anisotropy:asymmetries} and \ref{sec:pole_vs_eq}). Only for favored inclinations, e.g. when
$\mathbf{\Omega}$ points directly towards us and the TS ring is
resolved (as in the Crab nebula), can the synchrotron emission from the TS
be used to make more definite statements. Again, RMHD simulations for
anisotropic winds and synchrotron emission maps are needed to break
the degeneracies.

Pulsar bow shocks probe the local small-scale structure of the
ISM. An ISM density gradient modifies the shape of
the bow shock substantially away from the \citet{Wilkin00} solution. Since H$\alpha$
emission is generated behind the BS, density
fluctuations with a length-scale $\ga R_0$ should be clearly visible. If
the gradient is nearly perpendicular to $\mathbf{v}_m$, we expect the
surface brightness of the nebula to be asymmetric as well, even
one-sided in extreme cases. In addition, the TS is tilted towards the
low density side, producing asymmetric synchrotron
emission. Steep density gradients with length-scale $\la R_0$ (i.e. walls) produce
characteristic kink-like features. For example, a wall consisting of
mainly ionized matter can account for the observed shape of the PWN associated with PSR J2124$-$3358
\citep{Chatterjee06}. If the pulsar encounters a truncated wall with a finite width, the
bow shock develops a peculiar shape: an egg-like head and a
divergent tail (as in figure \ref{fig:res:denswallsaandsb}). This shape
is strikingly similar to the H$\alpha$ emission observed from PSR
B0740$-$28 \citep{Jones02}. Multi-epoch observations of the Guitar
nebula \citep{Chatt04} reveal time-dependent behaviour of this PWN
that corresponds to the overall situation shown in figure
\ref{fig:res:denswallseries} (although not to the time between the
snapshots in figure \ref{fig:res:denswallseries}, of course, which are
separated $\sim 10^2$ yr). When the pulsar runs into the
higher density  region, the wind is confined more strongly and the BS is
pushed closer to the pulsar, explaining the narrow head
and diverging tail. Furthermore, we note that the Guitar nebula
is significantly brighter near the strongly confined head. This
can be explained in terms of the increased density of the shocked ISM,
if we assume that the high density region carries a neutral fraction
comparable to the low density region. On the other hand, our
simulations do not exhibit the rear
shock seen in the Guitar nebula, which could be due to a
higher $\eta$ than considered in models SA and SB.

Summarizing our results, we conclude that the anisotropy of the wind
momentum flux alone cannot explain odd bow shock
morphologies. Instead, it is necessary to take into account external
effects, like ISM density gradients or walls. Conversely, because the
shape of the bow shock is degenerate with respect to several
parameters, it is difficult to infer the angular distribution of the
wind momentum flux from the H$\alpha$ radiation observed. Highly
resolved radio and X-ray observations, combined with synchrotron emission maps
from RMHD simulations, may break these degeneracies in the
future. Finally, we caution the reader that the results presented here
explore a very limited portion of the (large) parameter space of the problem.

\section*{Acknowledgements}
The software used in this work was in part developed by the
DOE-supported ASC/Alliance Center for Astrophysical Thermonuclear
Flashes at the University of Chicago. M.V., S.C. and
B.M.G. acknowledge the support of NASA through Chandra grant GO5-6075X
and LISA grant NAG5-13032. We also thank Ben Karsz for
assistance with aspects of the visualization.

\bibliography{bspaper.bib}

\begin{thebibliography}{}

\bibitem[\protect\citeauthoryear{{Arons}}{{Arons}}{2004}]{Arons04}
{Arons} J.,  2004, Advances in Space Research, 33, 466

\bibitem[\protect\citeauthoryear{{Bell}, {Bailes}, {Manchester}, {Weisberg} \&
  {Lyne}}{{Bell} et~al.}{1995}]{Bell95}
{Bell} J.~F.,  {Bailes} M.,  {Manchester} R.~N.,  {Weisberg} J.~M.,    {Lyne}
  A.~G.,  1995, \apjl, 440, L81

\bibitem[\protect\citeauthoryear{{Bogovalov}}{{Bogovalov}}{1999}]{Bogovalov99}
{Bogovalov} S.~V.,  1999, \aap, 349, 1017

\bibitem[\protect\citeauthoryear{{Bogovalov}, {Chechetkin}, {Koldoba} \&
  {Ustyugova}}{{Bogovalov} et~al.}{2005}]{Bogovalov05}
{Bogovalov} S.~V.,  {Chechetkin} V.~M.,  {Koldoba} A.~V.,    {Ustyugova} G.~V.,
   2005, \mnras, 358, 705

\bibitem[\protect\citeauthoryear{{Bucciantini}}{{Bucciantini}}{2002}]{Bucc02}
{Bucciantini} N.,  2002, \aap, 387, 1066

\bibitem[\protect\citeauthoryear{{Bucciantini}}{{Bucciantini}}{2006}]{Bucc06}
{Bucciantini} N.,  2006, astro-ph/0608258

\bibitem[\protect\citeauthoryear{{Bucciantini}, {Amato} \& {Del
  Zanna}}{{Bucciantini} et~al.}{2005}]{Bucc05}
{Bucciantini} N.,  {Amato} E.,    {Del Zanna} L.,  2005, \aap, 434, 189

\bibitem[\protect\citeauthoryear{{Bucciantini} \& {Bandiera}}{{Bucciantini} \&
  {Bandiera}}{2001}]{Bucc01}
{Bucciantini} N.,  {Bandiera} R.,  2001, \aap, 375, 1032

\bibitem[\protect\citeauthoryear{{Chatterjee} \& {Cordes}}{{Chatterjee} \&
  {Cordes}}{2002}]{Chatt02}
{Chatterjee} S.,  {Cordes} J.~M.,  2002, \apj, 575, 407

\bibitem[\protect\citeauthoryear{{Chatterjee} \& {Cordes}}{{Chatterjee} \&
  {Cordes}}{2004}]{Chatt04}
{Chatterjee} S.,  {Cordes} J.~M.,  2004, \apjl, 600, L51

\bibitem[\protect\citeauthoryear{{Chatterjee} et~al.,}{{Chatterjee}
  et~al.}{2006}]{Chatterjee06}
{Chatterjee} S.,  et~al., 2006, {in preparation}

\bibitem[\protect\citeauthoryear{{Chatterjee}, {Vlemmings}, {Brisken}, {Lazio},
  {Cordes}, {Goss}, {Thorsett}, {Fomalont}, {Lyne} \& {Kramer}}{{Chatterjee}
  et~al.}{2005}]{Chatt05}
{Chatterjee} S.,  {Vlemmings} W.~H.~T.,  {Brisken} W.~F.,  {Lazio} T.~J.~W.,
  {Cordes} J.~M.,  {Goss} W.~M.,  {Thorsett} S.~E.,  {Fomalont} E.~B.,  {Lyne}
  A.~G.,    {Kramer} M.,  2005, \apjl, 630, L61

\bibitem[\protect\citeauthoryear{{Chevalier}, {Kirshner} \&
  {Raymond}}{{Chevalier} et~al.}{1980}]{Chev80}
{Chevalier} R.~A.,  {Kirshner} R.~P.,    {Raymond} J.~C.,  1980, \apj, 235, 186

\bibitem[\protect\citeauthoryear{{Chevalier} \& {Raymond}}{{Chevalier} \&
  {Raymond}}{1978}]{Chev78}
{Chevalier} R.~A.,  {Raymond} J.~C.,  1978, \apjl, 225, L27

\bibitem[\protect\citeauthoryear{{Cordes}, {Romani} \& {Lundgren}}{{Cordes}
  et~al.}{1993}]{Cordes93}
{Cordes} J.~M.,  {Romani} R.~W.,    {Lundgren} S.~C.,  1993, \nat, 362, 133

\bibitem[\protect\citeauthoryear{{Coroniti}}{{Coroniti}}{1990}]{Coroniti90}
{Coroniti} F.~V.,  1990, \apj, 349, 538

\bibitem[\protect\citeauthoryear{{Cox} \& {Tucker}}{{Cox} \&
  {Tucker}}{1969}]{Cox69}
{Cox} D.~P.,  {Tucker} W.~H.,  1969, \apj, 157, 1157

\bibitem[\protect\citeauthoryear{{Del Zanna}, {Amato} \& {Bucciantini}}{{Del
  Zanna} et~al.}{2004}]{DelZanna04}
{Del Zanna} L.,  {Amato} E.,    {Bucciantini} N.,  2004, \aap, 421, 1063

\bibitem[\protect\citeauthoryear{{Deshpande}}{{Deshpande}}{2000}]{Deshpande00}
{Deshpande} A.~A.,  2000, \mnras, 317, 199

\bibitem[\protect\citeauthoryear{{Fryxell} et~al.,}{{Fryxell}
  et~al.}{2000}]{Fryxell00}
{Fryxell} B.,  et~al., 2000, \apjs, 131, 273

\bibitem[\protect\citeauthoryear{{Gaensler}, {Arons}, {Kaspi}, {Pivovaroff},
  {Kawai} \& {Tamura}}{{Gaensler} et~al.}{2002}]{Gaensler02a}
{Gaensler} B.~M.,  {Arons} J.,  {Kaspi} V.~M.,  {Pivovaroff} M.~J.,  {Kawai}
  N.,    {Tamura} K.,  2002, \apj, 569, 878

\bibitem[\protect\citeauthoryear{{Gaensler}, {Jones} \& {Stappers}}{{Gaensler}
  et~al.}{2002}]{Gaensler02}
{Gaensler} B.~M.,  {Jones} D.~H.,    {Stappers} B.~W.,  2002, \apjl, 580, L137

\bibitem[\protect\citeauthoryear{{Gaensler} \& {Slane}}{{Gaensler} \&
  {Slane}}{2006}]{Gaensler06a}
{Gaensler} B.~M.,  {Slane} P.~O.,  2006, {Annu. Rev. of Astron. \& Astrophys.},
  44, 17

\bibitem[\protect\citeauthoryear{{Gaensler}, {van der Swaluw}, {Camilo},
  {Kaspi}, {Baganoff}, {Yusef-Zadeh} \& {Manchester}}{{Gaensler}
  et~al.}{2004}]{Gaensler04}
{Gaensler} B.~M.,  {van der Swaluw} E.,  {Camilo} F.,  {Kaspi} V.~M.,
  {Baganoff} F.~K.,  {Yusef-Zadeh} F.,    {Manchester} R.~N.,  2004, \apj, 616,
  383

\bibitem[\protect\citeauthoryear{{Ghavamian}, {Raymond}, {Smith} \&
  {Hartigan}}{{Ghavamian} et~al.}{2001}]{Ghav01}
{Ghavamian} P.,  {Raymond} J.,  {Smith} R.~C.,    {Hartigan} P.,  2001, \apj,
  547, 995

\bibitem[\protect\citeauthoryear{{Goldreich} \& {Julian}}{{Goldreich} \&
  {Julian}}{1969}]{Goldreich69}
{Goldreich} P.,  {Julian} W.~H.,  1969, \apj, 157, 869

\bibitem[\protect\citeauthoryear{{Helfand}, {Gotthelf} \& {Halpern}}{{Helfand}
  et~al.}{2001}]{Helfand01}
{Helfand} D.~J.,  {Gotthelf} E.~V.,    {Halpern} J.~P.,  2001, \apj, 556, 380

\bibitem[\protect\citeauthoryear{{Hobbs}, {Lorimer}, {Lyne} \&
  {Kramer}}{{Hobbs} et~al.}{2005}]{Hobbs05}
{Hobbs} G.,  {Lorimer} D.~R.,  {Lyne} A.~G.,    {Kramer} M.,  2005, \mnras,
  360, 974

\bibitem[\protect\citeauthoryear{{Jones}, {Stappers} \& {Gaensler}}{{Jones}
  et~al.}{2002}]{Jones02}
{Jones} D.~H.,  {Stappers} B.~W.,    {Gaensler} B.~M.,  2002, \aap, 389, L1

\bibitem[\protect\citeauthoryear{{Kaspi}, {Roberts} \& {Harding}}{{Kaspi}
  et~al.}{2004}]{Kaspi04}
{Kaspi} V.~M.,  {Roberts} M.~S.~E.,    {Harding} A.~K.,  2004, {to appear in:
  {\it Compact Stellar X-ray Sources}, ed: Lewin, W.H.G. \& van der Klis, M.
  astro-ph/0402136}

\bibitem[\protect\citeauthoryear{{Kennel} \& {Coroniti}}{{Kennel} \&
  {Coroniti}}{1984a}]{Kennel84}
{Kennel} C.~F.,  {Coroniti} F.~V.,  1984a, \apj, 283, 694

\bibitem[\protect\citeauthoryear{{Kennel} \& {Coroniti}}{{Kennel} \&
  {Coroniti}}{1984b}]{Kennel84b}
{Kennel} C.~F.,  {Coroniti} F.~V.,  1984b, \apj, 283, 710

\bibitem[\protect\citeauthoryear{{Komissarov} \& {Lyubarsky}}{{Komissarov} \&
  {Lyubarsky}}{2003}]{Komissarov03}
{Komissarov} S.~S.,  {Lyubarsky} Y.~E.,  2003, \mnras, 344, L93

\bibitem[\protect\citeauthoryear{{Komissarov} \& {Lyubarsky}}{{Komissarov} \&
  {Lyubarsky}}{2004}]{Komissarov04}
{Komissarov} S.~S.,  {Lyubarsky} Y.~E.,  2004, \mnras, 349, 779

\bibitem[\protect\citeauthoryear{{Kulkarni} \& {Hester}}{{Kulkarni} \&
  {Hester}}{1988}]{Kulkarni88}
{Kulkarni} S.~R.,  {Hester} J.~J.,  1988, \nat, 335, 801

\bibitem[\protect\citeauthoryear{{Loehner}}{{Loehner}}{1987}]{Loehner87}
{Loehner} R.,  1987, {Comp. Meth. App. Mech. Eng.}, 61, 323

\bibitem[\protect\citeauthoryear{{Luo}, {McCray} \& {Mac Low}}{{Luo}
  et~al.}{1990}]{Luo90}
{Luo} D.,  {McCray} R.,    {Mac Low} M.-M.,  1990, \apj, 362, 267

\bibitem[\protect\citeauthoryear{{Melatos}}{{Melatos}}{1997}]{Melatos97}
{Melatos} A.,  1997, \mnras, 288, 1049

\bibitem[\protect\citeauthoryear{{Melatos}}{{Melatos}}{2002}]{Melatos02}
{Melatos} A.,  2002, in {Slane} P.~O.,  {Gaensler} B.~M.,  eds, ASP Conf. Ser.
  271: Neutron Stars in Supernova Remnants {Theory of Plerions}.
pp 115--+

\bibitem[\protect\citeauthoryear{{Melatos}, {Johnston} \& {Melrose}}{{Melatos}
  et~al.}{1995}]{Melatos95}
{Melatos} A.,  {Johnston} S.,    {Melrose} D.~B.,  1995, \mnras, 275, 381

\bibitem[\protect\citeauthoryear{{Melatos} \& {Melrose}}{{Melatos} \&
  {Melrose}}{1996}]{Melatos96}
{Melatos} A.,  {Melrose} D.~B.,  1996, \mnras, 279, 1168

\bibitem[\protect\citeauthoryear{{Melatos}, {Scheltus}, {Whiting},
  {Eikenberry}, {Romani}, {Rigaut}, {Spitkovsky}, {Arons} \& {Payne}}{{Melatos}
  et~al.}{2005}]{Melatos05}
{Melatos} A.,  {Scheltus} D.,  {Whiting} M.~T.,  {Eikenberry} S.~S.,  {Romani}
  R.~W.,  {Rigaut} F.,  {Spitkovsky} A.,  {Arons} J.,    {Payne} D.~J.~B.,
  2005, \apj, 633, 931

\bibitem[\protect\citeauthoryear{{Naz{\'e}}, {Chu}, {Guerrero}, {Oey},
  {Gruendl} \& {Smith}}{{Naz{\'e}} et~al.}{2002}]{Naze02}
{Naz{\'e}} Y.,  {Chu} Y.-H.,  {Guerrero} M.~A.,  {Oey} M.~S.,  {Gruendl} R.~A.,
     {Smith} R.~C.,  2002, \aj, 124, 3325

\bibitem[\protect\citeauthoryear{{Olson}, {MacNeice}, {Fryxell}, {Ricker},
  {Timmes} \& {Zingale}}{{Olson} et~al.}{1999}]{MacNeice99}
{Olson} K.~M.,  {MacNeice} P.,  {Fryxell} B.,  {Ricker} P.,  {Timmes} F.~X.,
  {Zingale} M.,  1999, Bulletin of the American Astronomical Society, 31, 1430

\bibitem[\protect\citeauthoryear{{Raymond}, {Cox} \& {Smith}}{{Raymond}
  et~al.}{1976}]{Raymond76}
{Raymond} J.~C.,  {Cox} D.~P.,    {Smith} B.~W.,  1976, \apj, 204, 290

\bibitem[\protect\citeauthoryear{{Roberts}, {Tam}, {Kaspi}, {Lyutikov},
  {Vasisht}, {Pivovaroff}, {Gotthelf} \& {Kawai}}{{Roberts}
  et~al.}{2003}]{Roberts03}
{Roberts} M.~S.~E.,  {Tam} C.~R.,  {Kaspi} V.~M.,  {Lyutikov} M.,  {Vasisht}
  G.,  {Pivovaroff} M.,  {Gotthelf} E.~V.,    {Kawai} N.,  2003, \apj, 588, 992

\bibitem[\protect\citeauthoryear{{Rosner}, {Tucker} \& {Vaiana}}{{Rosner}
  et~al.}{1978}]{Rosner78}
{Rosner} R.,  {Tucker} W.~H.,    {Vaiana} G.~S.,  1978, \apj, 220, 643

\bibitem[\protect\citeauthoryear{{Spitkovsky} \& {Arons}}{{Spitkovsky} \&
  {Arons}}{1999}]{Spitkovsky99}
{Spitkovsky} A.,  {Arons} J.,  1999, Bulletin of the American Astronomical
  Society, 31, 1417

\bibitem[\protect\citeauthoryear{{Spitkovsky} \& {Arons}}{{Spitkovsky} \&
  {Arons}}{2004}]{Spitkovsky04}
{Spitkovsky} A.,  {Arons} J.,  2004, \apj, 603, 669

\bibitem[\protect\citeauthoryear{{Stappers}, {Gaensler}, {Kaspi}, {van der
  Klis} \& {Lewin}}{{Stappers} et~al.}{2003}]{Stappers03}
{Stappers} B.~W.,  {Gaensler} B.~M.,  {Kaspi} V.~M.,  {van der Klis} M.,
  {Lewin} W.~H.~G.,  2003, Science, 299, 1372

\bibitem[\protect\citeauthoryear{{van der Swaluw}, {Achterberg}, {Gallant},
  {Downes} \& {Keppens}}{{van der Swaluw} et~al.}{2003}]{Swaluw03}
{van der Swaluw} E.,  {Achterberg} A.,  {Gallant} Y.~A.,  {Downes} T.~P.,
  {Keppens} R.,  2003, \aap, 397, 913

\bibitem[\protect\citeauthoryear{{van Kerkwijk} \& {Kulkarni}}{{van Kerkwijk}
  \& {Kulkarni}}{2001}]{Kerkwijk01}
{van Kerkwijk} M.~H.,  {Kulkarni} S.~R.,  2001, \aap, 380, 221

\bibitem[\protect\citeauthoryear{{Wilkin}}{{Wilkin}}{1996}]{Wilkin96}
{Wilkin} F.~P.,  1996, \apjl, 459, L31+

\bibitem[\protect\citeauthoryear{{Wilkin}}{{Wilkin}}{2000}]{Wilkin00}
{Wilkin} F.~P.,  2000, \apj, 532, 400

\end{thebibliography}

\appendix

\section{Analytic bow shock formulas in the thin-shell limit}
\label{sec:app:analytic}

\citet{Wilkin96} showed that the shape of the CD when the pulsar wind
momentum flux is isotropic can be written as:
\begin{equation}
  \label{eq:res:wilkin_simple}
  R(\theta)=R_0 \csc \theta [3(1-\theta \cot \theta)]^{1/2}
\end{equation}
in the limit of a thin shock. Here, $R(\theta)$ gives the distance
from the pulsar position to the bowshock,  $\theta$ denotes the
colatitude measured relative to the orientation of the symmetry axis
(see section \ref{sec:anisotropy}), and $R_0$ is the standoff distance
defined in \eref{res:standoff}.

The formalism was extended to an anisotropic pulsar wind with $p=2$,
by \citet{Wilkin00}, who found
\begin{eqnarray}
  \label{eq:res:wilkin}
    R(\theta)&=&R_0 \csc \theta' \{
        3(1-\theta' \cot \theta')[c_0 + \frac{c_2}{4}(3v^2+w^2)]
      \nonumber \\
        & & + \frac{3 c_2}{4} [ (v^2-w^2)\sin^2\theta'+v w
          (2\theta'-\sin 2\theta') ] \}^{1/2}, \nonumber \\
        & &
\end{eqnarray}
with $v=\sin \phi' \sin \lambda$, $w=\cos \lambda$, and
$c_0=1-c_2/3$, if the momentum flux is of the form
$\rho_w(\theta)=\rho_{0,w}(c_0+c_2 \cos^p \theta)$ as in equation
\eeref{aniso:momflux}. $\theta'$ and $\phi'$ are the usual spherical
coordinates with respect to the $z$ axis, such that
\begin{equation}
  \label{eq:app:trans}
  \cos \theta =  v \sin \theta' +  w \cos \theta'.
\end{equation}
Note that this notation differs from \citet{Wilkin00}: his
$\theta_\ast$ corresponds to our $\theta$, while his $\theta$ and
$\phi$ correspond to our $\theta'$ and $\phi'$ respectively.

We derive a formula for $R(\theta)$ in the case $p=4$ by applying
the formalism of \citet{Wilkin00}. The case $p=4$ is relevant to a
collimated jet \citep{Chatterjee06} or an
extended vacuum dipole (see appendix
\ref{sec:app:momflux}). Inside the wind cavity, the momentum flux can
be written as \citep{Wilkin00}
\begin{equation}
  \rho_w(r, \theta) v_w^2=\frac{\dot{E}}{2 \pi r^2 v_w} g_w(\theta),
\end{equation}
with
\begin{equation}
  \label{eq:app:momflux}
  g_w(\theta)=c_0+c_2 \cos^4 \theta.
\end{equation}
Here, the overall energy loss $\dot{E}$ is related to the simulation
parameters $\rho_{0,w}$ and $r_w$ through
$\dot{E}=2 \pi r_w^2 \rho_{0,w} v_w^3$.
The normalisation\footnote{Although this normalisation is not
necessary for the analytic solution, it is required for
\eref{aniso:momflux} to hold. If it is not fulfilled, the definition
of the scale factor $R_0$ changes accordingly.}
\begin{equation}
  \int_0^{2 \pi} \mathrm{d}\phi \int_0^{\pi} \mathrm{d}\theta \,
  \sin \theta \, g_w(\theta)=4\pi
\end{equation}
requires $c_0=1-c_2/5$ for $p=4$. The incident momentum flux from the
wind on the shell can then be written as
\begin{equation}
  \mathbf{\Phi}_w(\theta', \phi')=\frac{\dot{E}}{2 \pi  v_w}
  \mathbf{G}_w(\theta', \phi').
\end{equation}
Here, $\mathbf{G}_w=G_{w, \varpi} \hat{\bvarpi}+G_{w,z} \hat{\mathbf{z}}$, where
the cylindrical coordinates $\varpi$ and $z$ are defined with respect to
the symmetry axis, is a dimensionless function given by
\begin{equation}
  \label{eq:app:momfluxint}
  \mathbf{G}_w = \int_0^{\theta'}  \mathrm{d}\theta'' \, g_w (\hat{\bvarpi} \sin
  \theta'' + \hat{\mathbf{z}} \cos \theta'') \sin \theta''.
\end{equation}
From \eeref{app:trans} and \eeref{app:momflux}, we can compute $g_w$
and hence $\mathbf{G}_w$ as a function of $\theta'$ and $\phi'$:
\begin{equation}
  \label{eq:app:momfluxtrans}
  g_w = c_0 + c_2 (\sin \theta' v + \cos \theta' w)^4.
\end{equation}
It can now be shown \citep{Wilkin00} that the bow shock shape is
described by
\begin{equation}
  \label{eq:app:bowshockshape}
  R(\theta', \phi') = R_0 \csc \theta' [ -6 (G_{w, \varpi} \cot
    \theta' - G_{w,z})]^{1/2}.
\end{equation}
Substituting \eeref{app:momfluxint} and \eeref{app:momfluxtrans} into
\eeref{app:bowshockshape} finally yields

\begin{eqnarray}
  \label{eq:app:wilkinp4}
  R & = &\frac{R_0}{8} \csc \theta' \{ -\csc \theta'
    \{ c_0[ 192\theta' \cos \theta' -  192\sin \theta']
   \\  & & +
    c_2[ vw^3
      ( 16\cos \theta' - 12\cos 3\theta' - 
        4\cos 5\theta' - 96\theta' \sin\theta' )
      \nonumber \\  & & +
     v^3w( 32\cos \theta' - 36\cos 3\theta' + 
        4\cos 5\theta' - 96\theta' \sin \theta' )
      \nonumber \\
   & &+ 
     v^2w^2( 144\theta' \cos \theta' - 168\sin\theta' + 
        18\sin 3\theta' - 6\sin 5\theta' ) \nonumber \\
   & & +
     v^4( 120\theta' \cos \theta' - 80\sin \theta' - 
        15\sin 3\theta' + \sin 5\theta' ) \nonumber \\
   & & +
     w^4( 24\theta' \cos \theta' - 56\sin \theta' + 
        9\sin 3\theta' + \sin 5\theta' )  ]\}
    \}^{1/2}. \nonumber
\end{eqnarray}

\begin{figure}
  \centering
  \includegraphics[width=41mm, keepaspectratio]{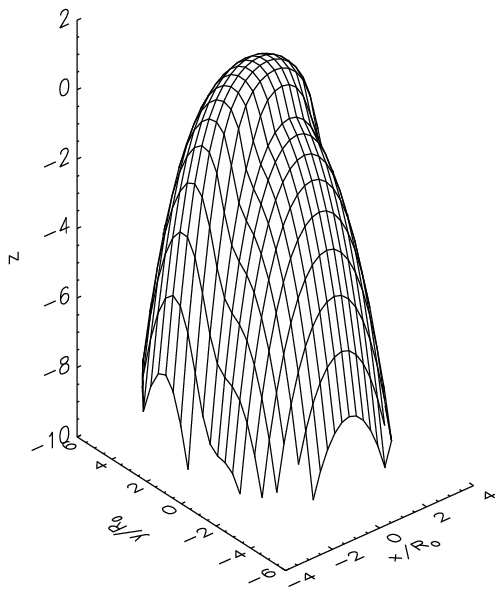}
  \includegraphics[width=41mm, keepaspectratio]{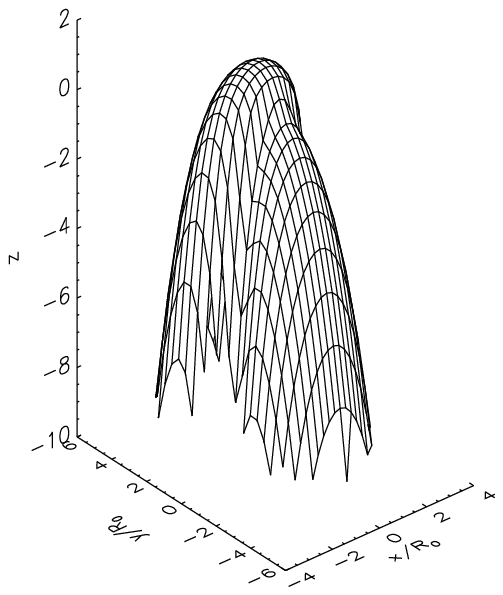}
  \caption{Comparison of the bow shock shape for $p=2$ (left) and
    $p=4$ (right), given $\lambda=\pi/4$, $c_0=0$ and $c_2=3$. We note
  the strong similarity between these solutions, the only difference
  being the slimmer shape for $p=4$.}
  \label{fig:app:shapes}
\end{figure}

\begin{figure}
  \centering
  \includegraphics[width=84mm, keepaspectratio]{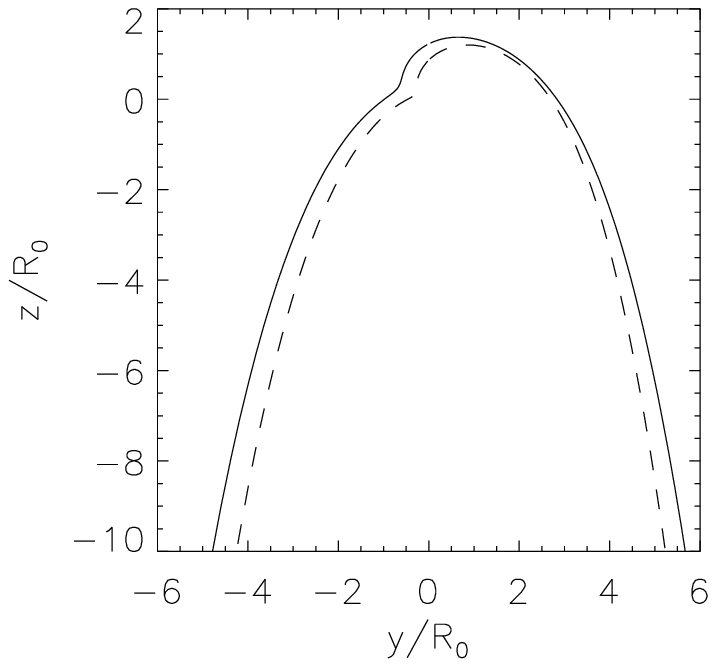}
  \caption{$y$-$z$ section of the bow shock for $p=2$ (solid
    curve) and $p=4$ (dashed curve). The strong similarity hampers
    efforts to distinguish between the two cases observationally.}
  \label{fig:app:shape2d}
\end{figure}

Figure \ref{fig:app:shapes} shows the two solutions \eeref{res:wilkin}
and \eeref{app:wilkinp4} for $\lambda=\pi/4$, $c_0=0$, and $c_2=3$. Although the shapes are nearly identical,
the $p=4$ surface is slimmer. The difference is seen more
clearly in figure \ref{fig:app:shape2d}, which shows the $y$-$z$
section of the bow shock. The $p=4$ case (dashed curve) touches the
$p=2$ shock (solid curve) at the latitude of maximum momentum flux, $\theta=\pi/4$
(upper-right corner). Both surfaces show a characteristic kink at
the latitude where the momentum flux is a minimum.

By a similar calculation, we can derive the shape of the BS for an
exponential density gradient perpendicular to $\mathbf{v}_m$
\citep{Wilkin00}. The result is
\begin{equation}
  \label{eq:res:wilkindens}
  R(\theta, \phi)=H \sec \phi \sin^{-1} \theta y_s(\theta, \phi),
\end{equation}
where $y_s$ is the solution of
\begin{equation}
  \frac{1}{y} \left[ y-2+\exp(-y)(y+2)\right] = \frac{1}{6} y_\mathrm{ax}^2
\end{equation}
and $y_\mathrm{ax}$ is the normalised standoff distance,
\begin{equation}
  y_\mathrm{ax}=\frac{R_0}{H}\cos\phi [3 (1-\theta \cot \theta)]^{1/2}.
\end{equation}

\section{Angular distribution of the momentum flux for an extended
  vacuum dipole}
\label{sec:app:momflux}
Although the electrodynamics of the pulsar magnetosphere and wind
remain unsolved, it is commonly assumed that the angular distribution of the
energy flux at the light cylinder is preserved in the
transition to a kinetic-energy-dominated outflow at the TS
\citep{DelZanna04}. In this appendix, we calculate the far-field Poynting flux for
point-like and extended dipoles rotating in vacuo.

The relevant electromagnetic far field components are given by the real
parts of the following expressions \citep{Melatos97}:
\begin{equation}
  \label{eq:farfield1}
  B_\theta=\frac{\mathrm{i}B_0}{2 x} \left[a(x_0)-b(x_0)\right] \sin
  \alpha \cos \theta \mathrm{e}^{\mathrm{i}(x+\eta)}+O(x^{-2}),
\end{equation}
\begin{equation}
  B_\phi=-\frac{B_0}{2x}\left[a(x_0) \cos 2\theta - b(x_0)\right] \sin
  \alpha \mathrm{e}^{\mathrm{i(x+\eta)}}+O(x^{-2}),
\end{equation}
\begin{equation}
  \label{eq:farfield3}
  (E_\theta, E_\phi)=c(B_\phi, -B_\theta).
\end{equation}
In \eeref{farfield1}--\eeref{farfield3}, $x=\omega r/c$ denotes a
normalized radial coordinate, $\omega$ is the pulsar's angular velocity, $x_0=\omega
r_0/c$ is the normalized effective radius of the corotating
magnetosphere, $\eta=\phi-\omega t$ is the phase of the
electromagnetic wave, and $B_0$ is the magnitude of the stellar
magnetic field at
the poles. Constants $a$ and $b$ are given by
\begin{equation}
  a=\frac{x_0^2}{x_0 h_2'(x_0)+h_2(x_0)}
\end{equation}
and
\begin{equation}
  b=\frac{x_0}{h_1(x_0)},
\end{equation}
where $h_1(x)$ and $h_2(x)$ are the first and second-order spherical Hankel
functions of the first kind. The $r$ component of the Poynting flux
can then be computed from
\begin{eqnarray}
  \mathbf{S}_r & = & \frac{1}{2 \mu_0} (\mathbf{E} \times
  \mathbf{B^*})_r \nonumber \\
  & = & S_0 ( |a-b|^2 \cos^2 \theta + |a \cos 2\theta-b|^2) \sin^2 \alpha
  \label{eq:poynting}
\end{eqnarray}
with $S_0=(c/2\mu_0) (B_0/2x)^2$.

We can simplify \eeref{poynting} using the explicit form of the Hankel functions,
\begin{equation}
  \label{eq:app:hankel1}
  h_1(x)=\mathrm{e}^{\mathrm{i}x}
  \left(-\frac{1}{x}-\frac{\mathrm{i}}{x^2} \right)
\end{equation}
and
\begin{equation}
  \label{eq:app:hankel2}
  h_2(x)=\mathrm{e}^{\mathrm{i}x}\left(\frac{\mathrm{i}}{x}
      -\frac{3}{x^2}-\frac{3 \mathrm{i}}{x^3} \right).
\end{equation}
In the case of a point dipole, we have $x_0 \ll 1$. Retaining the
leading order terms, we obtain
\begin{equation}
  a=-\frac{\mathrm{i}}{6} x_0^5 \mathrm{e}^{-\mathrm{i}x_0},
\end{equation}
\begin{equation}
  b=-\mathrm{i}x_0^3 \mathrm{e}^{-\mathrm{i}x_0},
\end{equation}
and hence
\begin{equation}
  \mathbf{S}_r =S_0 x_0^6 (\cos^2 \theta +1) \sin^2 \alpha ,
\end{equation}
which reduces to the case $p=2$, $c_0=c_2=1$ in \citet{Wilkin00}.

We also calculate the Poynting flux for an extended dipole, as proposed by
\citet{Melatos97}. For the special case $x_0=1$, an explicit calculation yields
\begin{equation}
  \mathbf{S}_r =\frac{S_0}{17} ( 2 \cos^4 \theta + 6 \cos^2 \theta +5)
   \sin^2 \alpha.
\end{equation}
\end{document}